\documentclass[iop]{emulateapj}

\usepackage{amsmath,amssymb,graphicx,natbib}

\newcommand{\kepler}{\textit{Kepler}}

\shorttitle{Multihabitable Systems}
\shortauthors{Steffen \& Li}

\begin{document}

\title{Dynamical considerations for life in multihabitable planetary systems}
\author{Jason H. Steffen\altaffilmark{1,2}}
\affil{University of Nevada, Las Vegas, Department of Physics and Astronomy, 4505 S. Maryland Pkwy., Las Vegas, NV 89154-4002 }

\author{Gongjie Li}
\affil{Harvard Smithsonian Center for Astrophysics, Institute for Theory and Computation, 60 Garden Street, Cambridge, MA 02138}

\altaffiltext{1}{Northwestern University, CIERA, 2131 Tech Drive, Evanston, IL 60208}
\altaffiltext{2}{Lindheimer Fellow}

\begin{abstract}
Inspired by the close-proximity pair of planets in the Kepler-36 system, we consider two effects that may have important ramifications for the development of life in similar systems where a pair of planets may reside entirely in the habitable zone of the hosting star.  Specifically, we run numerical simulations to determine whether strong, resonant (or non-resonant) planet-planet interactions can cause large variations in planet obliquity---thereby inducing large variations in climate.  We also determine whether or not resonant interactions affect the rate of lithopanspermia between the planet pair---which could facilitate the growth and maintenance of life on both planets.  We find that first-order resonances do not cause larger obliquity variations compared with non-resonant cases.  We also find that resonant interactions are not a primary consideration in lithopanspermia.  Lithopanspermia is enhanced significantly as the planet orbits come closer together---reaching nearly the same rate as ejected material falling back to the surface of the originating planet (assuming that the ejected material makes it out to the location of our initial conditions).  Thus, in both cases our results indicate that close-proximity planet pairs in multihabitable systems are conducive to life in the system.
\end{abstract}



\section{Introduction}
NASA's \kepler\ mission has discovered several thousand candidate exoplanet systems including many hundreds with multiple planets \citep{Borucki:2010,Steffen:2010,Batalha:2013,Burke:2014,Rowe:2015,Mullaly:2015}.  Among the many interesting systems identified by this mission is Kepler-36, which has two planets orbiting near the 7:6 mean-motion resonance (MMR) \citep{Carter:2012}.  These two planets have orbital periods of 13.8 and 16.2 days, have masses of 4.5 and 8.1 M$_\oplus$, and radii of 1.5 and 3.7 R$_\oplus$ for the inner and outer planets respectively.  It is striking to note that these two planets (with densities differing by a factor of 8) orbit at distances that differ by only 10\%.  And, while Kepler-36 is among the most extreme cases currently known, several other pairs of planets lie near first-order MMRs with similarly high indices such as the 6:5 and 5:4.

Given the context of a mission designed to measure the number of habitable planets in the galaxy and the alien nature of many of the systems that \kepler\ has found (e.g., circumbinary planets \citep{Doyle:2011,Welsh:2012b}, planetary systems orbiting each component of a binary pair Kepler-132 \citep{Lissauer:2014}, systems with planets near chains of MMR such as Kepler-80 \citep{Xie:2013} and Kepler-223 \citep{Lissauer:2011b}, and these systems with pairs of planets in strongly interacting (and for the case of Kepler-36 manifestly chaotic orbits \citep{Deck:2012}) it is not much of a stretch to consider systems with multiple planets that orbit in the habitable zone of their host stars, including pairs in or near MMR.

Here we consider two issues that have been raised regarding life on Earth but in the context of these Kepler-36-inspired multihabitable systems.  First, whether or not close planetary orbits or resonant interactions cause any significant variations in planet obliquities---the angle between the planet's spin and orbital axes.  Obliquity variation plays a major role in the modulation of climate, since it determines the latitudinal distribution of solar radiation.  For the case of Mars (an ocean-free, atmosphere-ice-regolith system), the obliquity changes would result in drastic variations of atmospheric pressure caused by runaway sublimation of CO$_2$ ice \citep{Toon80, Fanale82, Pollack82, Francois90, Nakamura03, Soto12}.

For Earth-like planets (planets partially covered by oceans) the change of climate depends on the specific land-sea distribution and on the position within the habitable zone around the star.  The ice ages on the Earth, for example, are closely associated with the variation in insolation at high latitudes, which depends on the orbital eccentricity and orientation of the spin axis according to the Milankovitch theory \citep[e.g.][]{Weertman76, Hays76, Imbrie82}.  While it is debatable whether the variation in obliquity truly renders a planet uninhabitable (obliquity variations may, in some cases, actually increase the habitability of a planet \citep{Armstrong:2014}, though civilizations that rely on agriculture may struggle), it is clear that the climate can change drastically as the obliquity varies \citep{Williams97, Chandler00, Jenkins00, Spiegel09}. 

The spin-axis dynamics of planets in the solar system has been extensively studied in the literature.  At present, the obliquity variation of the Earth is regular and only undergoes small oscillations between $22.1^\circ$ and $24.5^\circ$ with a 41000 year period \citep[e.g.][]{Vernekar72, Laskar93b}. Without the Moon, the obliquity of the hypothetical Earth is chaotic, but is constrained between $0 - 45^\circ$ over billion year timescales \citep{Laskar93a, Lissauer12, Li14a}---though \citet{Lissauer12} showed some conditions where the obliquity of the Earth can be stable in the absence of the Moon.  The Earth's obliquity remained stable as the Moon moved outward before Late Heavy Bombardment \citep{Li14b}, yet Martian obliquity is thought to have been chaotic throughout the solar system's lifetime \citep{Ward73, Touma93, Laskar93a, Brasser11}.  If strong planet-planet interactions in multihabitable systems cause or preclude large obliquity variations, then the development of intelligent life would be hampered or aided by its effects on climate.

The second issue we address, perhaps more rich in consequences, is to understand how biological material might be exchanged between planets in a multihabitable system through the process of lithopanspermia (hereafter simply ``panspermia'').  Panspermia in the solar system has been studied at some length \citep[e.g.,][see this last reference for a good historical review]{Melosh:1993,Gladman:1996,Worth:2013}. And, from observations of meteorites, some hypothesize that a fraction of life on Earth may have originated on Mars (which may have supported life sooner than the Earth \citep[e.g.,][]{McKay:1996}).  The survival of life on collision ejecta, both at the time of collision \citep{Melosh:1998} and over the subsequent, extended trajectory was studied by \citet{Mileikowsky:2000}, who found that successful transfer (especially from Mars to Earth) was highly probable.

Two Earth-like planets in Kepler-36-like orbits would likely have a much greater opportunity to exchange such material than the terrestrial planets in the solar system.  The planets would subtend over 25$\times$ the solid angle at conjunction than the Earth does from Mars and the relative velocities of the planets, and hence the ejecta particles, could be much less.  This scenario allows more of the ejected particles (especially with low relative velocities) to successfully make the trip between planets.  The close proximity of the two planets reduces the reliance on the effects of secular resonances to successfully transfer particles \citep{Dones:1999}.  (For the solar system, secular resonances excite orbital eccentricities of the ejected particles and facilitate their transfer over large distances.)

\begin{figure}
\includegraphics[width=0.49\textwidth]{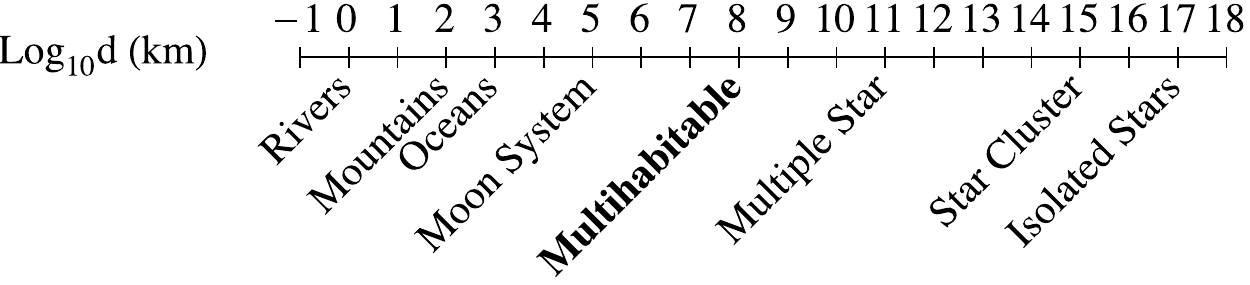}
\caption{Logarithm of various length scales over which life could be seeded via panspermia.  This paper is primarily concerned with multihabitable systems.\label{lengthscales}}
\end{figure}

Consider the scales over which biological material may be transmitted via collision ejecta (see Figure \ref{lengthscales}), on the shortest scales, individual habitable planets may have barriers such as oceans and mountain ranges that could be traversed by ejecta from a planet falling back to onto the same planet, ``auto-panspermia'' (which has been shown as a viable means of seeding life across the Earth \citep{Wells:2003}).  The distances involved would be $\sim 0.1 \rightarrow 10^4$km.  Binary planets or a single planet with multiple habitable moons are on a somewhat larger scale, perhaps $10^5 \rightarrow 10^7$km.  On the largest scales one can (and has) considered interstellar panspermia within a star cluster or within the galaxy ($10^{13} \rightarrow 10^{19}$km) \citep{Melosh:2003,Belbruno:2012}.  Between these extremes lie habitable planets orbiting different components of a stellar binary ($10^{10} \rightarrow 10^{12}$km).  And, finally, multihabitable systems ($10^8 \rightarrow 10^9$km), which we consider here.

This paper is organized as follows.  In the next section we briefly discuss the properties of the model systems we consider.  Section \ref{obliquity} presents the evolution of planet obliquities in the model systems.  Section \ref{transport} shows the results of our panspermia simulations.  We discuss some of the implications of our work in section \ref{discussion} and give concluding remarks in section \ref{conclusion}.

\section{Planetary system models}\label{models}

For this investigation, we construct a sample of systems in first-order MMR along with systems that have nearly identical period ratios but that are not in MMR.  The systems in MMR were created following the methods of \citet{Lee02, Batygin10}.  Specifically, we set the planets slightly outside of their resonant location, and integrate the system using a Burlisch-Stoer integrator.  We evolve the planets under semi-major axis and eccentricity damping in addition to the Newtonian N-body interaction.

In these simulations, the central star has a mass of one solar mass, and the planets each have masses of $1.0\times 10^{-6}$ M$_\odot$ (about 1/3 of an Earth mass).  The planet masses were chosen arbitrarily, and our results should not depend strongly on this quantity.  We ran a large number of simulations with different semi-major axis and eccentricity decay rates, and selected the final state of the systems where the planets are in MMR at the end of the simulations.  The corresponding non-resonant configurations are obtained by setting the longitude of pericenter, longitude of ascending node, and the mean anomaly to zero for both planets.  A third set of non-resonant initial conditions was also chosen for the purposes of differentiating the effects due to resonance and the effects due to initial conditions.  All of our initial conditions are given in Table \ref{initcond} in the Appendix.

\section{Obliquity evolution}\label{obliquity}

The obliquity variation in a planet arises largely as a consequence of the underlying resonant structure \citep{Laskar96}.  Specifically, the spin-axis of the planet may exhibit complex behavior if its precession resonates with the inclination variation.  The former is controlled primarily by the stellar torques, whereas the latter is forced by planet-planet interactions.  When the precession frequency of the planetary spin axis from stellar torque coincides with the frequency of inclination variation from planetary interactions, the resonance occurs.

\begin{figure}
\includegraphics[width=0.49\textwidth]{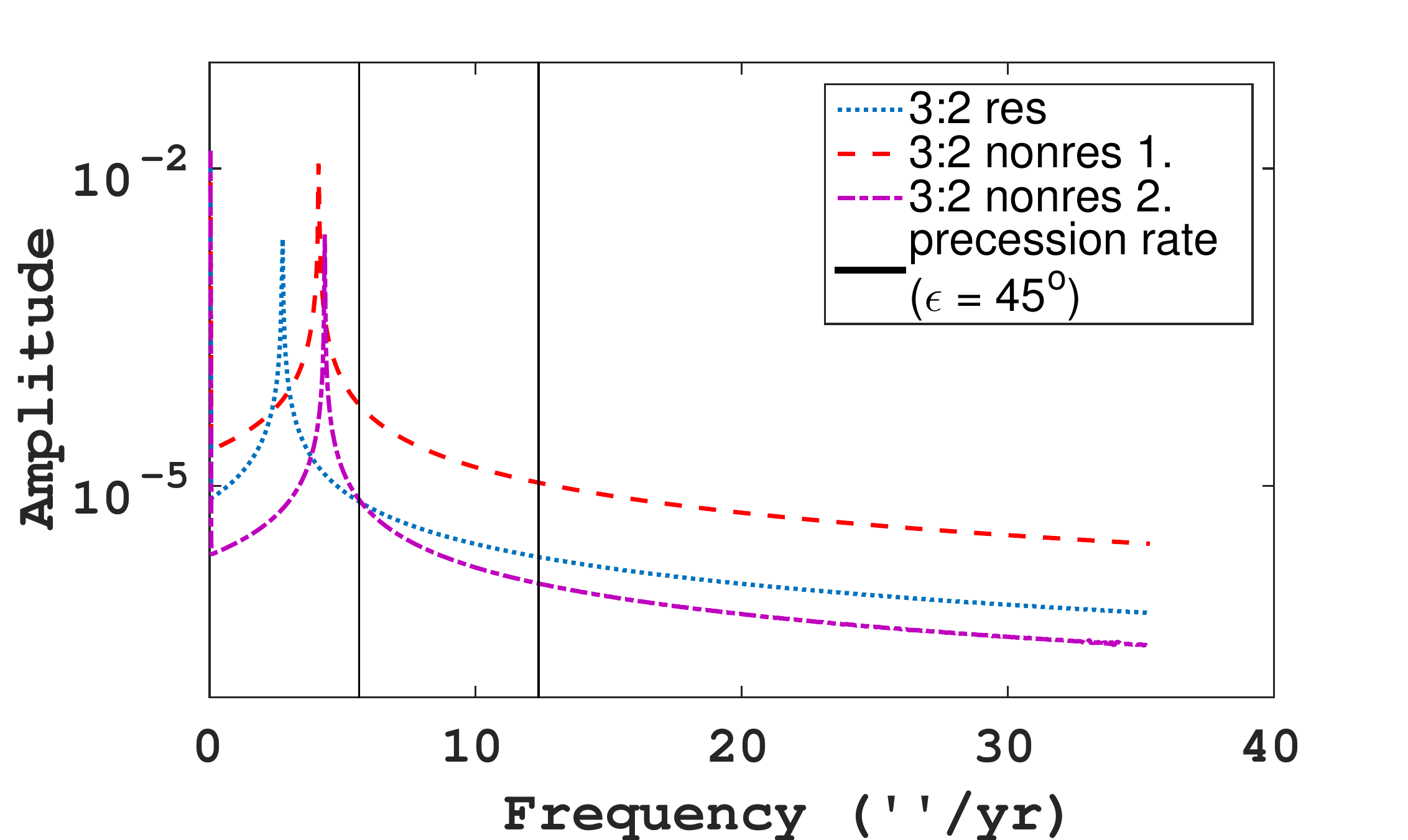}
\caption{The Fourier spectrum of ($i e^{\hat{i}\Omega}$) for the 3:2 resonance (where $\hat{i} \equiv \sqrt{-1}$). The blue dotted line represents the case when the two planets are in resonance, the red dashed line represents the case when the two planets are not in resonance and the angle variables (longitude of pericenter, longitude of ascending node, and mean anomaly) are set to 0, 120, and 240 degrees respectively, and the purple line represents the case when the angle variables are set to zero.  The maximum high-amplitude frequencies are approximately the same for the three cases.  The black lines show the precession of the inner and the outer planets assuming the planets have Earth-like properties.\label{f:freq}}
\end{figure}

For our systems, the stellar torque for both the resonant and the non-resonant systems are essentially the same, since the two systems only differ in the orbital orientation.  To calculate the frequency of the orbital inclination variation for the systems, we integrate the orbital elements numerically using the HYBRID integrator in the MERCURY package \citep{Chambers:1999}.  The Fourier spectrum of the planetary orbital parameter ($i e^{\hat{i}\Omega}$), where $\hat{i} \equiv \sqrt{-1}$, $i$ is the orbital inclination, and $\Omega$ is the longitude of ascending node, are shown in Figure \ref{f:freq}.  The amplitudes and frequencies of the inclination variations are approximately the same.  This fact indicates that first-order mean motion resonances do not affect obliquity variations---otherwise the resonant case would be different from the nonresonant ones.  The black lines show the precession of the inner and the outer planets assuming the planets have Earth-like properties.

\begin{figure}
\includegraphics[width=0.49\textwidth]{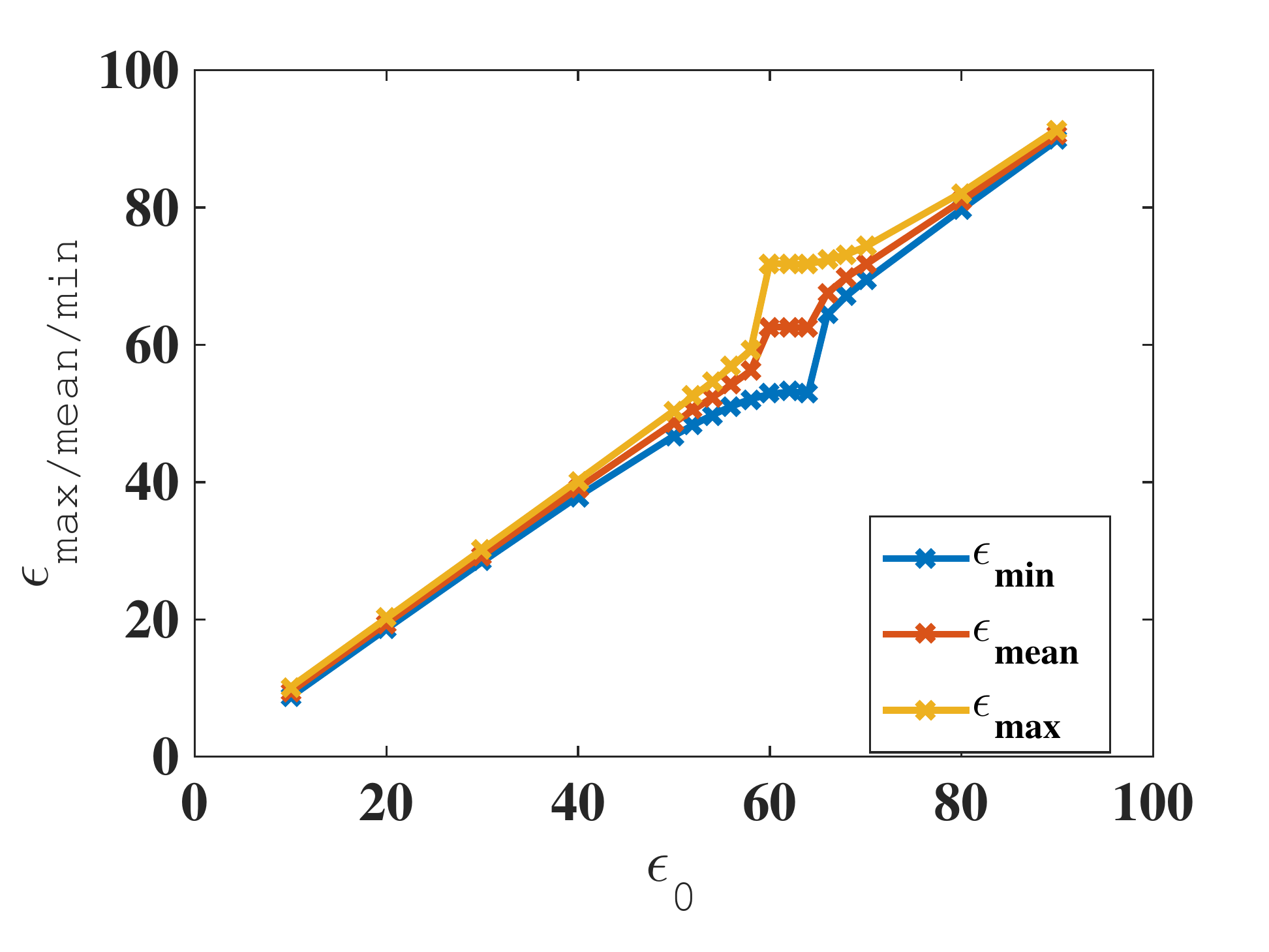}
\caption{The obliquity variation for the 3:2 resonance in 30 Myr. The precession frequency matches that of the inclination variation at $\epsilon \sim 62^\circ$, and causes the obliquity variation at around $62^\circ$.\label{f:obi}}
\end{figure}

\begin{figure*}
\includegraphics[width=\textwidth]{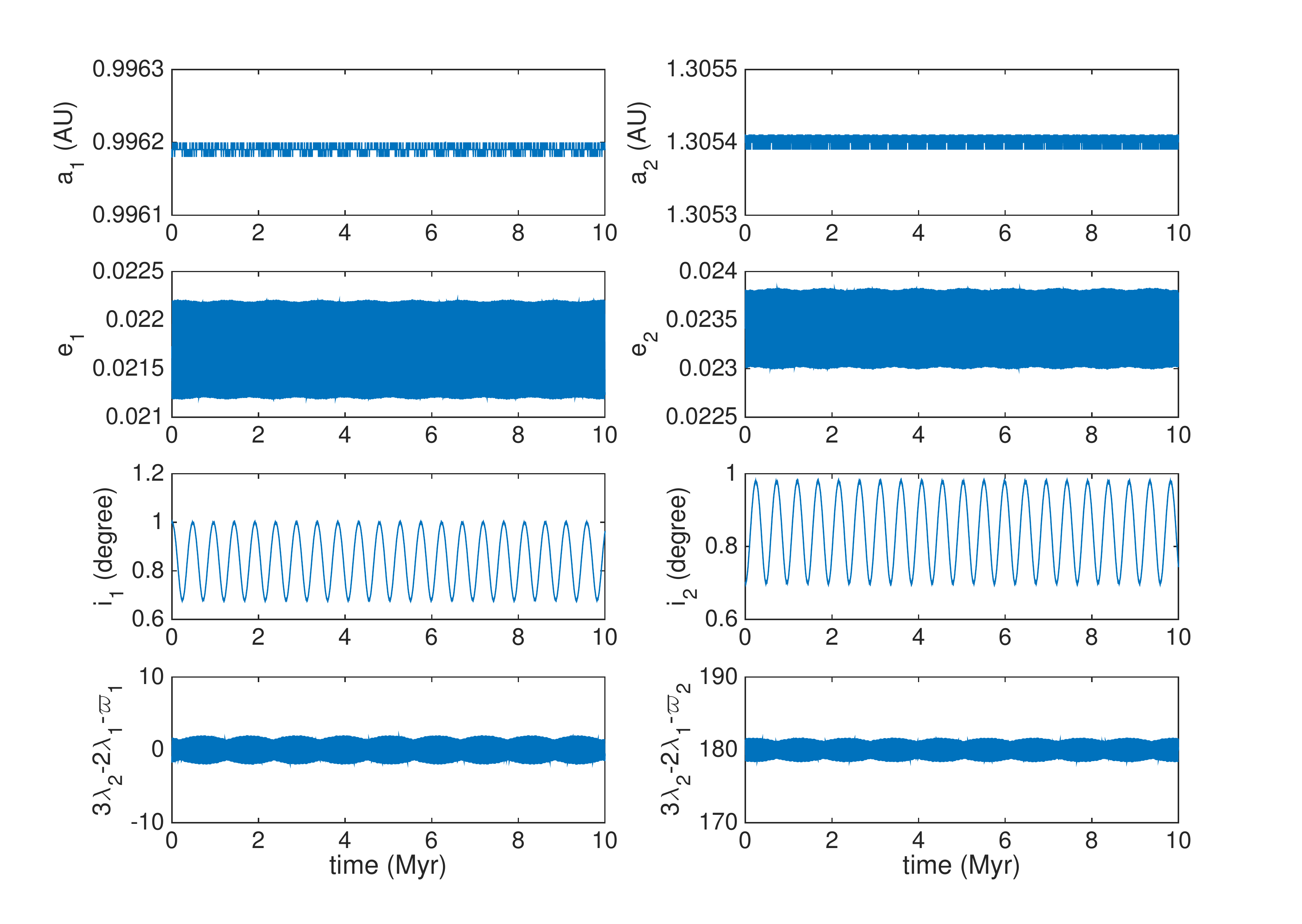}
\caption{\textbf{The evolution of the orbital elements for the 3:2 resonance case.  Here $\lambda$ is the longitude of the planet in its orbit, $\varpi$ is the longitude of pericenter, $e$ is the orbital eccentricity, $a$ is the orbital semi-major axis, and $i$ is the orbital inclination.  The libration of these quantities demonstrates the resonance behavior.} \label{f:res32}}
\end{figure*}

To illustrate how the obliquity variation occurs, we calculate the obliquity as a function of time following the Hamiltonian that describes its evolution \citep[e.g.][]{Colombo66, Kinoshita:1972, Laskar93a, Touma93, deSurgy97,Armstrong:2004}.  (The detailed equations are given in the Appendix.)  We set the planetary system configuration to be the case of ``3:2 nonresonant I'' (the red-dashed line) shown in Figure \ref{f:freq} and described in section \ref{resnonres}, the dynamical ellipticity of the planet is set to that of Earth ($E_d = 3 \times 10^{-3}$), and the rotation period of the planet is $12$ hrs.  This rotation period is somewhat arbitrary as we only require that the precession rate of the inner planet's spin axis, $\alpha \cos{(\epsilon)}$ (where $\alpha$ is the precession coefficient and $\epsilon$ is the obliquity---see the appendix for details), matches the inclination variation frequency when $\epsilon \sim 62^{\circ}$.  Figure \ref{f:obi} shows the maximum, the minimum and the mean of the obliquity variation as a function of the initial obliquity for $m_1$ in $\sim 10$ Myr.  This scenario, where the inclination and precession rates coincide, corresponds to the region where the obliquity of the planet varies with moderate amplitude.  The evolution of the orbital elements for the resonant case is shown in Figure \ref{f:res32}.

We briefly consider two additional scenarios relating to obliquity variation in a planetary system, though a detailed study of these scenarios lies beyond the scope of this work.  First, we consider the case where the Earth, embedded in the solar system, is replaced by a pair of near resonant planets with equal masses in order to determine whether or not the size of the obliquity variations differ significantly form those of a moonless Earth.  Second, we consider the case where a habitable, Mars-mass planet lives in a system with a near-resonant planet pair.

To illustrate the obliquity variation for systems involving closely separated planets, we compared the variation of a moonless Earth in the solar system with an analogous solar system where the Earth is replaced with two planets of mass $m_1=m_2=1\times10^{-6}$ $M_{\odot}$ located at $a_1=0.974$ AU, and $a_2=1.068$ AU (with period ratio $6/7$).  Different from the case of the solar system, one of the frequencies of the inclination modes peaks at $\sim 15.5''/$yr, and the mode at $\sim 20''/$ yr is lower in amplitude for $m_1$ and $m_2$.

\begin{figure}
\includegraphics[width=3in]{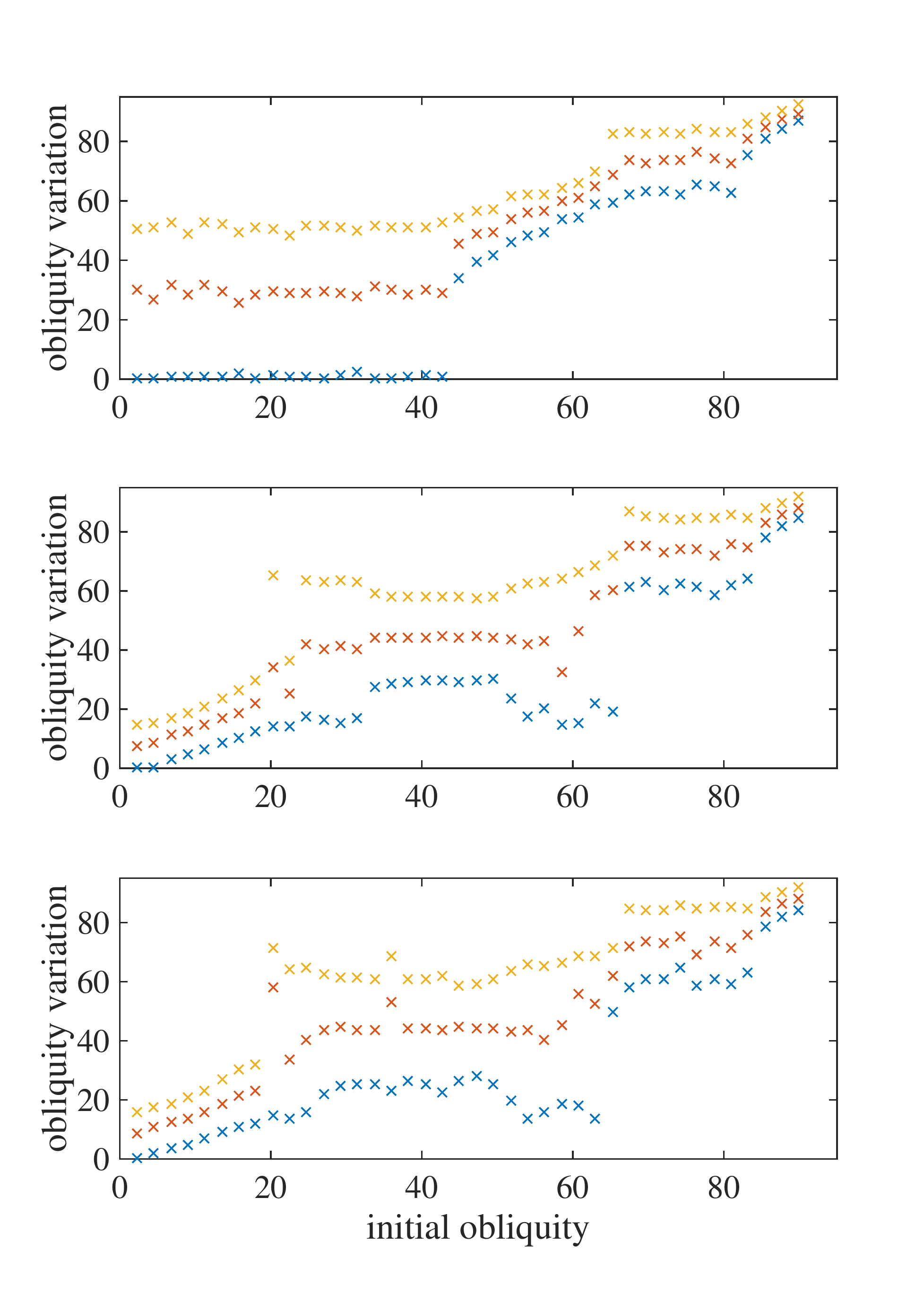}
\caption{The obliquity variation of planets. The upper panel shows the maximum, minimum and the mean obliquity in 20 Myr of a moonless Earth versus the initial obliquity, and the lower panels show those of a planet in the system where the Earth is substituted by two closely separated planets. Specifically, the middle panel shows the obliquity variation of the inner substituted planet ($m_1$), and the lower panel shows that of the outer substituted planet ($m_2$).  The precession coefficient $\alpha$ is set to be $21.24''/$yr. At this value, the obliquity variation of the planet pair is larger than that of the moonless Earth when the initial obliquity is $\sim 0-20$ degrees and weaker when it is $\sim 45-65$ degrees.\label{f:obearth}}
\end{figure}

The obliquity variation of the planets depends upon their precession coefficients, which we assume to be the same for both planets.  The obliquity variations of this ``split Earth'' scenario are shown in figure \ref{f:obearth} and are compared with variations of the true Earth.  When the precession coefficient of the planets are $21.24''/$yr, the variation is stronger for the true Earth where the initial obliquity is in the range of $\sim 0-20$ degrees, and is weaker when in the range of $\sim 45-65$ degrees.  The variation is quite similar between the true and split Earth scenarios when the initial obliquity is between $\sim20-45$ degrees.

Using the same split-Earth modification to the solar system, we also tested the variation of the obliquity of Mars.  This investigation addresses the issue of a habitable planet residing in the same system as a close pair of planets.  In this case we found that the obliquity variations of Mars were not strongly affected by the near resonant pair.

These results show that climate change due to obliquity variations for multihabitable systems embedded in the solar system would be comparable to, but not larger than, the variations of a moonless Earth.  The same is true for habitable planets living in a system with a close-proximity pair where the planet-planet interactions do not affect obliquity variations in any material way.  Only in relatively rare instances, where orbital parameters are somewhat fine-tuned (so that the timescales of inclination variation and spin precession coincide), does a planet's obliquity vary significantly.

\citet{Brasser:2014} showed that spin-orbit resonances could exist in exoplanet systems.  And, systems may be driven to such couplings through some damping mechanism (e.g., tides).  However, the scenarios that more readily produce spin-orbit couplings would likely be closer to the host star than the typical habitable zone (e.g., near the orbit of Mercury rather than Earth).  Thus, barring other influences, we can expect stable climates for planets in systems with resonant or near-resonant planet pairs at the same rate as stable climates in systems without them.

\section{Lithopanspermia}\label{transport}

Now we consider panspermia in these systems---the transmission of collision ejecta containing biological material from one planet to the other.  As we study panspermia in this context, there are a few dynamical points worth making.  First, since collision ejecta comes from a highly localized source and since the particles are collisionless (widely dispersed) following the initial impact, the available phase space for the particles is restricted (by Liouville's theorem).  Consequently, collision ejecta is dynamically very cold---having a very low, and ever decreasing, velocity dispersion.  The stream of debris will therefore be concentrated in phase-space sheets that fold and twist as the system evolves \citep[see, e.g., ][]{Sikivie:1998}---forming caustics or cold flows or, the term we will use, ``caustic flows''.

The fact that the debris evolves in this manner means that when one piece of debris strikes a planet, there is excess probability that additional pieces will also strike the planet and within a relatively short amount of time.  A consequence of this effect would be seen in the distribution of intercollision times.  If the debris were randomly spread in the vicinity of the planets then the distribution of intercollision times would be approximately exponential (the collisions being a Poisson process).  Here, the deviations from Poisson behavior would come primarily from long-term reduction of the ejecta population through collisions or ejections.  On the other hand, the collision time intervals from caustic flows would have an excess of collisions separated by short times.  An overall Poisson behavior would be visible only over long times as the planets pass through the various flows.  We investigate these effects below.

Another issue to consider is the strong dependence of the results of any panspermia calculation on the chosen distribution of initial velocities.  With relative ease one can choose a velocity distribution where all of the particles are ejected from the system---leaving only a single epoch for dynamical encounters between the destination planet and the ejecta (effectively eliminating the probability of a successful transfer).  However, if the typical ejection velocities put the particles on orbits with periods comparable to the destination planet (or slightly more or less than the escape velocity from the originating planet) then successful transfers are more likely to occur.

\subsection{Setup}

For our panspermia simulations we consider both when the inner planet and the outer planet act as the source.  We take each planetary system and evolve it over one synodic time to select a set of 10 initial configurations of the system that are roughly equidistant in the relative longitudes of the two planets.  We then rotate these initial configurations so that the longitude of the source planet is always equal to zero.  This transformation gives a common substellar point on the source planet for analysis purposes.

For each initial configuration we divide the surface of the source planet into 768 locations using Hierarchical Equal Area isoLatitude Pixelisation \citep[HEALPix,][]{Gorski:2005}.  Each location is the source of three ejecta particles with randomly assigned velocities directed radially outward from the surface.  The starting position of each particle is located one Hill radius from the planet above the geometric center of each pixel.  The choice to start at one Hill radius is to eliminate potential numerical artifacts related to the planet surface.  Thus, for each planetary system and for each source planet in that system we had 10 initial planet locations, each with 768 ejecta starting locations, each with 3 initial velocities giving 23,040 particles per simulation.

For each velocity, we choose a uniformly distributed number between 0.5 and 0.5 $v_\star/v_p$ where $v_p$ is the escape velocity from the planet (beginning at one Hill radius) and $v_\star$ is the escape velocity from the star at the orbital distance of the inner planet (all set to 1 AU).  This random number $\mathcal{R}$ is then substituted into the formula:
\begin{equation}
v = v_p \sqrt{1+\mathcal{R}^2}.
\label{vdistequation}
\end{equation}
This particular form for the function that defines the velocities is quite arbitrary and was motivated by the desire to have roughly the appropriate energy scale for the escaped particles in the system (between just escaping from the planet and just escaping from the star).  When added to the orbital motion of the planet, some of the highest velocity particles along the direction of orbital motion escape the system, especially when the outer planet is the source.  Throughout the simulations the collision ejecta are treated as test particles and did not affect the orbits of the planets.  We integrated each system for 10 million years using MERCURY---employing the HYBRID integrator.

\subsection{Resonant and non-resonant planet pairs}\label{resnonres}

The first test to identify any differences between resonant and nonresonant systems.  We ran simulations as described above for both resonant and nonresonant planet pairs and for both inner and outer planets as the source.  The four resonances we consider are the 7:6, 6:5, 4:3, and 3:2.  For these systems the time dependence of the successful transfers were very similar, generally giving of order 1000 events.  However, we noticed that the nonresonant systems consistently had a $\sim 20$\% higher success rate regardless of planet source or MMR.

We ran a set of 5 simulations for the 6:5 nonresonant case in order to estimate the variations that we could expect from different initial conditions for the velocities of the ejected particles.  The results show that the variation over the course of the whole integration is typical of Poisson fluctuations in the successful transfers (statistical variations of only a few percent given our $\sim 1000$ successes).  This fact implies that the difference between the resonant and nonresonant success rates are quite significant---generally between 2.5 and 6 sigma for the different MMRs.

This result may have been due to the fact that the angle variables (longitude of pericenter, longitude of ascending node, and mean anomaly) were all set to zero---giving a highly specialized initial condition.  Consequently, we ran additional simulations for the 4:3 and 3:2 MMRs with nonresonant planet pairs where the longitudes of pericenter and ascending node, and the mean anomaly for the inner planet were set to 0, 120$^\circ$, and 240$^\circ$ respectively (these are the ``I'' cases as shown in Figure \ref{f:freq}).  For these new nonresonant cases the number of successful transfers were either similar to, or somewhat less than, the resonant cases (and significantly below the original nonresonant case).

That the transfer rate from the second set of nonresonant simulations differed significantly from the original set shows that the planetary initial conditions play a strong role in panspermia (imagine two planets with very different orbital inclinations).  The variations that arise from changing a few orbital angles are comparable to or larger than the variations between resonant and nonresonant systems.  This fact suggests that the primary cause of the observed differences between the resonant and nonresonant simulations may be due primarily to the initial conditions of the planets rather than their resonant behavior, or lack thereof.  A definitive statement in this regard, since its effects must be very small ($\lesssim 1$\%), would require a study that lies beyond our scope.  Nevertheless, since the ejecta themselves are unlikely to be in resonance with the destination planet, it would not be surprising that the resonance behavior of the planets is of little consequence.



\subsection{Variations with Longitude}

In our simulations we found that the successful transfers depended somewhat on the initial longitude of the particles.  When the outer planet was the source, more successfully transferred particles were ejected opposite the direction of motion.  When the inner planet was the source, more successfully transferred particles were ejected from the substellar and antistellar points.  This longitudinal variation should depend upon the distribution of initial velocities.  Consider the inner planet source, in the extreme where the particles barely escape the planet, only the particles in the direction of motion would have sufficient energy to reach the orbit of the outer planet.  In the other extreme, with high velocity ejecta, only particles that directly face the outer planet at the time of the collision would transfer successfully.

To demonstrate this effect, we ran a suite of simulations using a different initial velocity distribution.  In this case the velocities were again assigned using Equation (\ref{vdistequation}) but with the random number being chosen from an exponential distribution with a scale parameter of $\lambda = 0.1 v_\star / v_p$.  We call this the ``restricted'' velocity distribution as it has a much smaller range of values for the assigned particle velocities.  The motivation for this distribution was simply to have a contrasting example.  Using these simulations we examine three observable consequences for successfully transferred particles and compare them with the results of the ``standard'' velocity distribution from before.  These observables are: 1) the longitudinal dependence of the successful transfers, 2) the transfer rate as a function of the separation of the planetary orbits, and 3) the inter-arrival time of the collisions (testing the Poisson nature of the collision rate).

\begin{figure}
\includegraphics[width=0.49\textwidth]{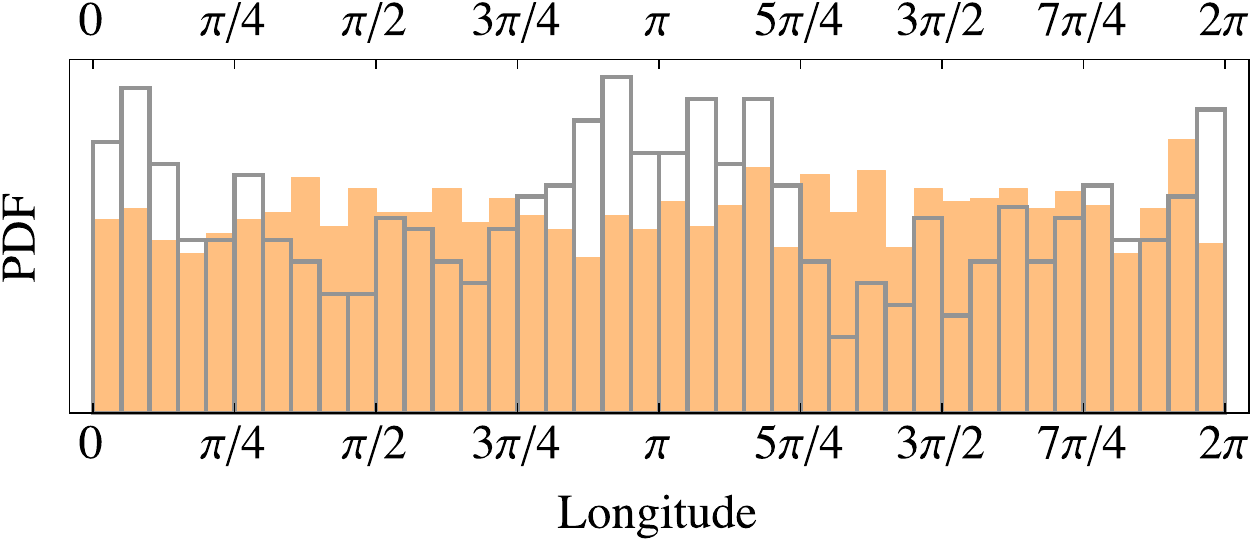}
\vskip0.1in
\includegraphics[width=0.49\textwidth]{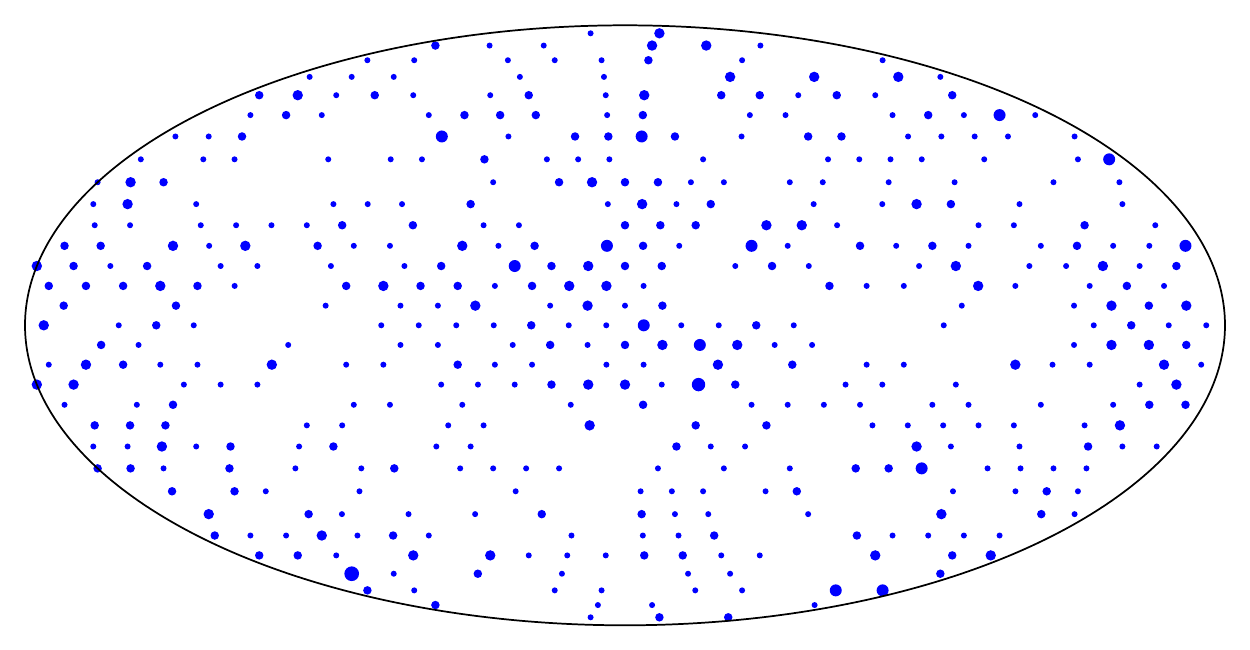}
\includegraphics[width=0.49\textwidth]{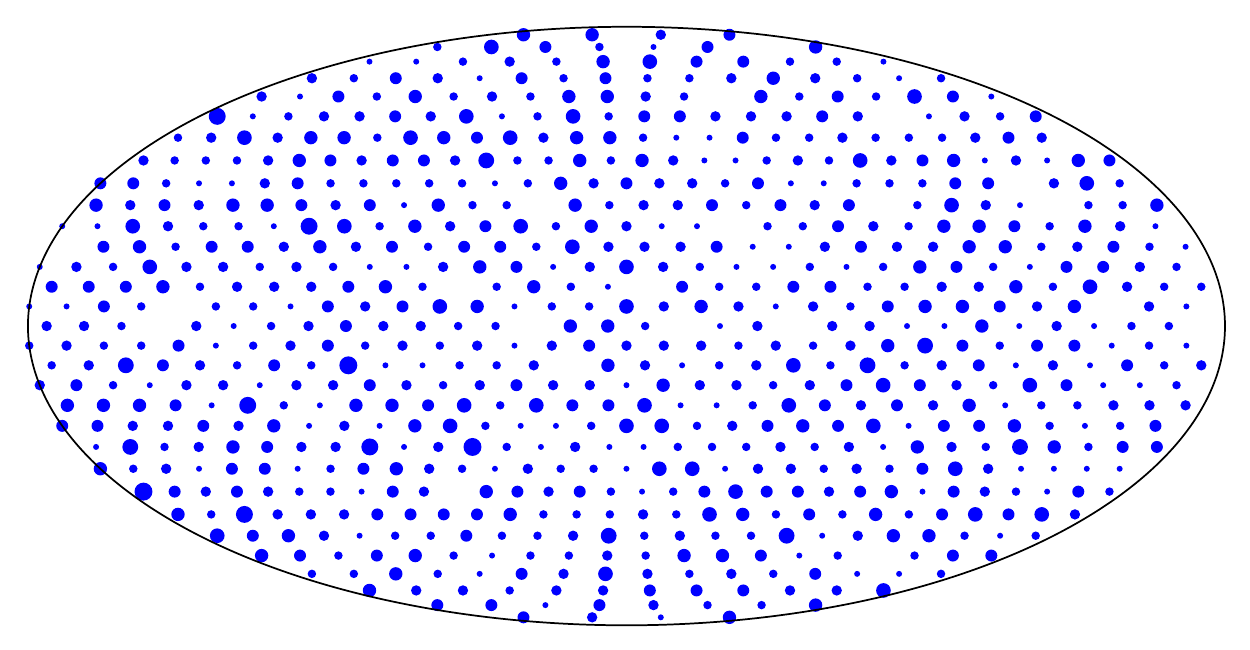}
\caption{Longitudinal (top) and spatial distribution (middle and bottom) of initial locations of ejecta from the inner planet for successful transfers from one planet to another near the 4:3 MMR.  The middle panel shows the initial locations for the standard velocity distribution and produce the outline histogram in the top panel.  The bottom panel shows the initial locations for the restricted velocity distribution and produce the solid histogram in the top panel.  The histograms show that the velocity distribution affects the longitudinal distribution of source points for successful transfers.  Here 0 corresponds to the sub-stellar point and $3 \pi /2$ is the direction of motion.  For the bottom panels, the areas of the circles are proportional to the number of successful transfers from that location (30 total are possible, but typically only one to a few are realized).}\label{hitpics}
\end{figure}

Figure \ref{hitpics} shows an example of the original spatial locations of the transferred particles for both the standard and the restricted velocity distributions.  The bottom two panels show the locations of the initial ejecta on the inner planet for the two velocity distributions.  The top panel, which is a histogram of the initial longitudes only, had a small random number added to the true longitude values.  The reason for this addition was to enable more reliable statistical tests---since the initial locations of the ejecta were discrete, statistical tests that rely on the empirical distribution (e.g., the Anderson-Darling test) are compromised because there are large jumps whenever multiple particles from the same longitude are considered.  These large jumps yield anomalously large differences between the empirical and comparison distributions.  The added random numbers were normally distributed with zero mean and a standard deviation of $2 \pi / 64$ (there are 32 initial locations around the equator of the source planet and this quantity is half of their separations).

Table \ref{pvalstable} shows the Anderson-Darling $p$-values comparing the distribution of the initial longitudes for the two initial velocity distributions both to a uniform distribution and to each other.  For all MMRs, and for both planet sources, a uniform distribution can not be excluded for the restricted velocity distribution.  For the standard velocity distribution, a uniform distribution is excluded with high confidence for most cases (there is some variation in these values depending upon the random numbers added to each longitude, but the changes do not affect the order of magnitude of the $p$-values).


\begin{table}
\caption{Anderson Darling $p$-values for the velocity distributions compared with a uniform distribution and with each other.\label{pvalstable}}
\begin{tabular}{cclll} \hline
MMR & Source & Standard & Restricted & S v. R \\ \hline
3:2 & 1 & 0.22 & 0.34 & 0.19 \\
3:2 & 2 & 0$^\star$ & 0.28 & 0 \\
4:3 & 1 & 0.03 & 0.23 & 0.02 \\
4:3 & 2 & 0 & 0.78 & 0 \\
6:5 & 1 & $3\times 10^{-6}$ & 0.35 & $7\times 10^{-5}$ \\
6:5 & 2 & 0 & 0.23 & 0 \\
7:6 & 1 & $2\times 10^{-5}$  & 0.73 & $8\times 10^{-4}$ \\
7:6 & 2 & 0 & 0.32 & 0 \\ \hline
\end{tabular}
\\
$^\star$ Zero values indicate that the $p$-value was less than machine precision.
\end{table}

\begin{figure}
\includegraphics[width=0.49\textwidth]{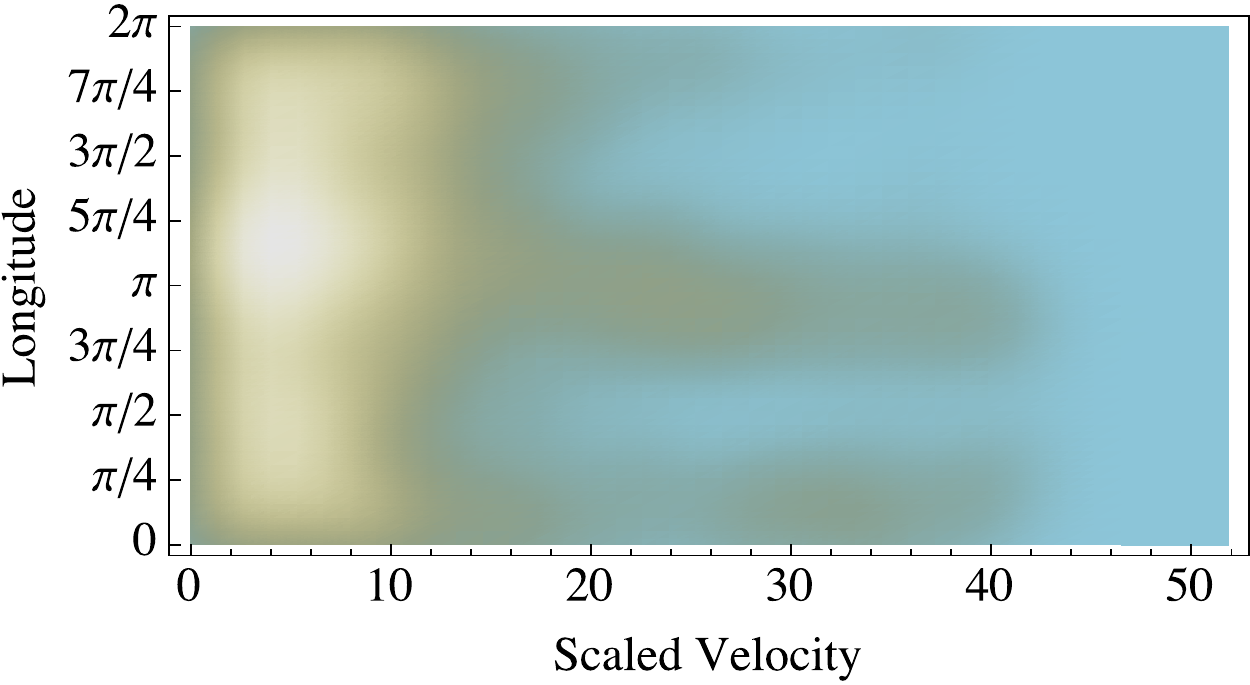}
\vskip0.1in
\includegraphics[width=0.49\textwidth]{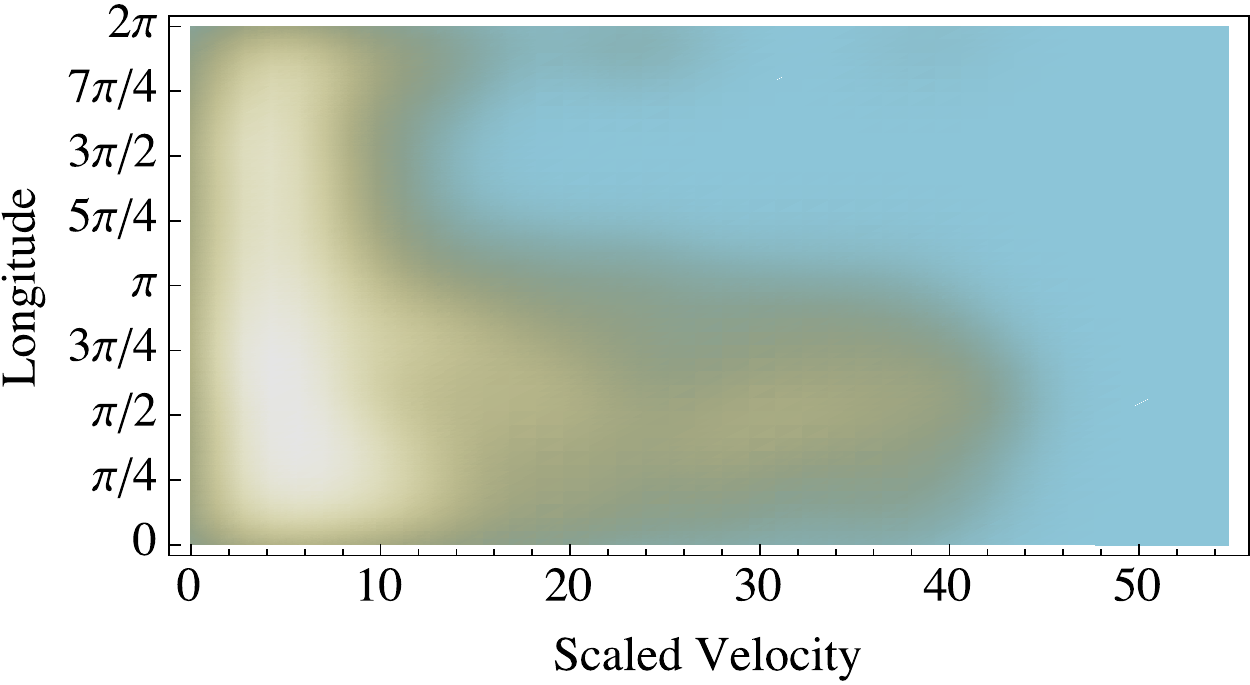}
\vskip0.1in
\center{\includegraphics[width=0.3\textwidth]{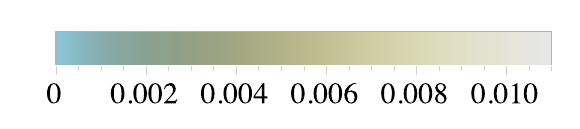}}
\caption{A smoothed density histogram of the fraction of successfully transferred particles as a function of the scaled velocity and longitude for the inner (top) and outer (bottom) planets for the standard velocity distribution.  Note that the range in velocities is larger at the substellar and antistellar points for the inner planet source and it is larger for the direction opposite the orbital motion for the outer planet.}\label{veldist}
\end{figure}

For the successful transfers using the standard velocity distribution, there is also a dependence on the velocity that varies with longitude.  Figure \ref{veldist} shows the distribution of the initial longitudes of the particles as a function of their their initial velocities for both the inner planet and the outer planet as source (the case shown is for a pair near the 4:3 MMR).  As one may expect, when the outer planet is the source, the particles simply need to lose sufficient energy to cross the inner planet orbit and the range of initial velocities is much larger away from the direction of motion than the range of velocities in the direction of motion.  A similar, somewhat more interesting, effect can be seen when the inner planet is the source where there is a wide range of successful initial velocities in the substellar and anti-stellar directions.  Similar results occur for other MMRs.  However, since the restricted velocity distribution is much more narrow than the standard distribution, the same structure does not appear.

In addition to the differences in longitudes of successful transfers between the two velocity distributions, we look at changes in the transfer rate that depend upon the difference in semi-major axis of the two planets.  The actual transfer rates are much higher for the restricted velocity distribution (by about a factor of three).  This difference is due in part to the more densely occupied phase space of the particles, there being far less spread in the velocities.  However, \citet{Gladman:1996} showed that the Martian meteorites on Earth also showed evidence for initial velocities near the escape velocity from Mars---a result consistent with our statement.

\begin{figure}
\includegraphics[width=0.49\textwidth]{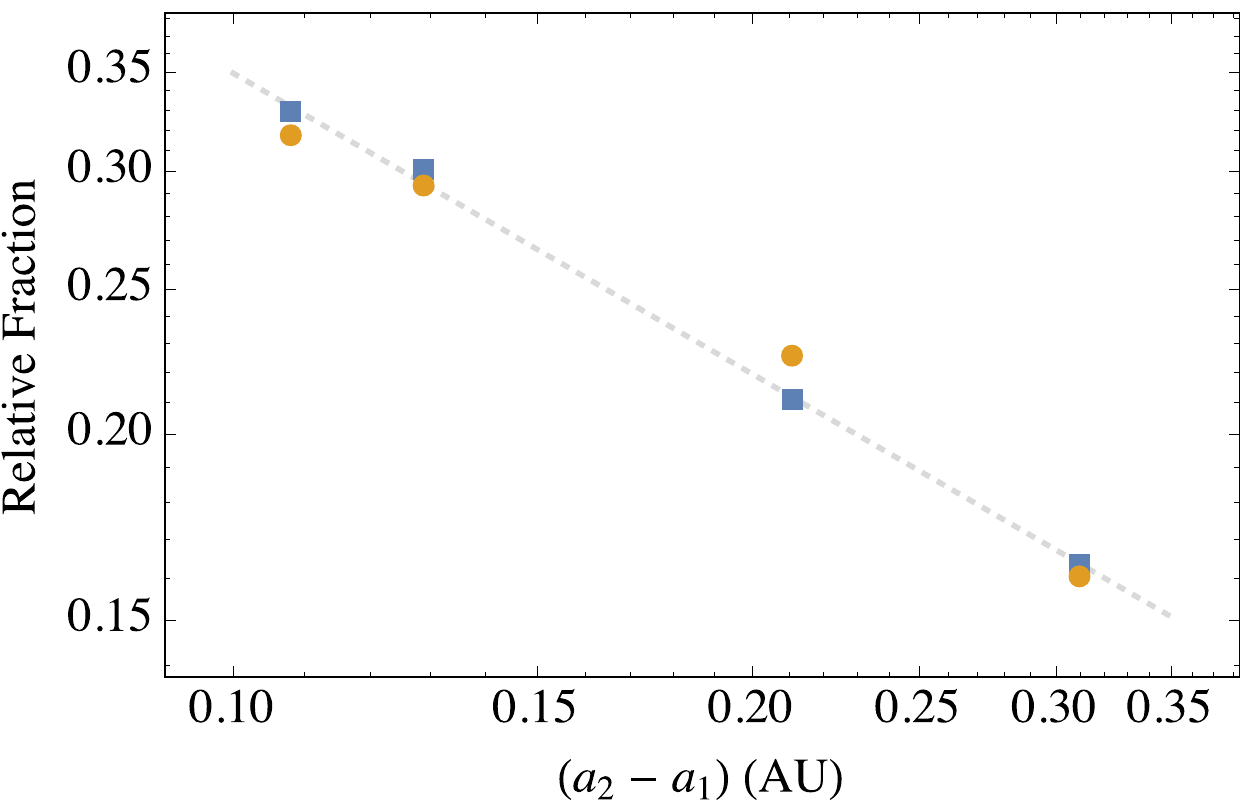}
\caption{The relative fraction of all successful transfers of collision ejecta ($N_{j\text{+}1:j}/N_\text{all}$), where $N$ is the number of successful transfers, from one planet to the other as a function of the difference between the orbital semi-major axes (the total of the $y$-values is unity for each case).  The blue squares correspond to the standard velocity distribution while the orange circles correspond to the restricted distribution.  The dotted line shows that for our two initial velocity distributions the relative fraction falls as $|a_2-a_1|^{-2/3}$ (only the normalization was allowed to float in a fit to the results of the standard distribution).  We believe the fact that these two examples are near this line to be a coincidence (see text).  Nevertheless, these results confirm the expectation that the closer the two planets are to each other, the more likely it is for material to transfer from one to the other.}\label{collisionfraction}
\end{figure}

Regardless of the overall rates, the relative rates for the different MMRs between these two velocity distributions are quite similar.  Figure \ref{collisionfraction} shows the relative fraction of successful transfers for both velocity distributions as a function of the difference in semi-major axes of the two planetary orbits.  (That is, for example, the number of successful transfers near the 7:6 or 6:5 MMR divided by the total number of successful transfers for all MMRs---$N_{j\text{+}1:j}/N_\text{all}$.)  Recall that the estimated fractional uncertainties in these rates is only a few percent and could not explain the observed behavior.  Power-law fits to the two sets of points both yield declining efficiency that scales approximately as $\sim (a_2 - a_1)^{-2/3}$ (shown for reference).  This apparent similarity between the results of the two velocity distributions is somewhat coincidental since one can construct limiting cases where there are no successful transfers at all (by having very small velocities) or where the rate falls as $(a_2 - a_1)^{-2}$ (by having very large ones).

\begin{figure}
\includegraphics[width=0.49\textwidth]{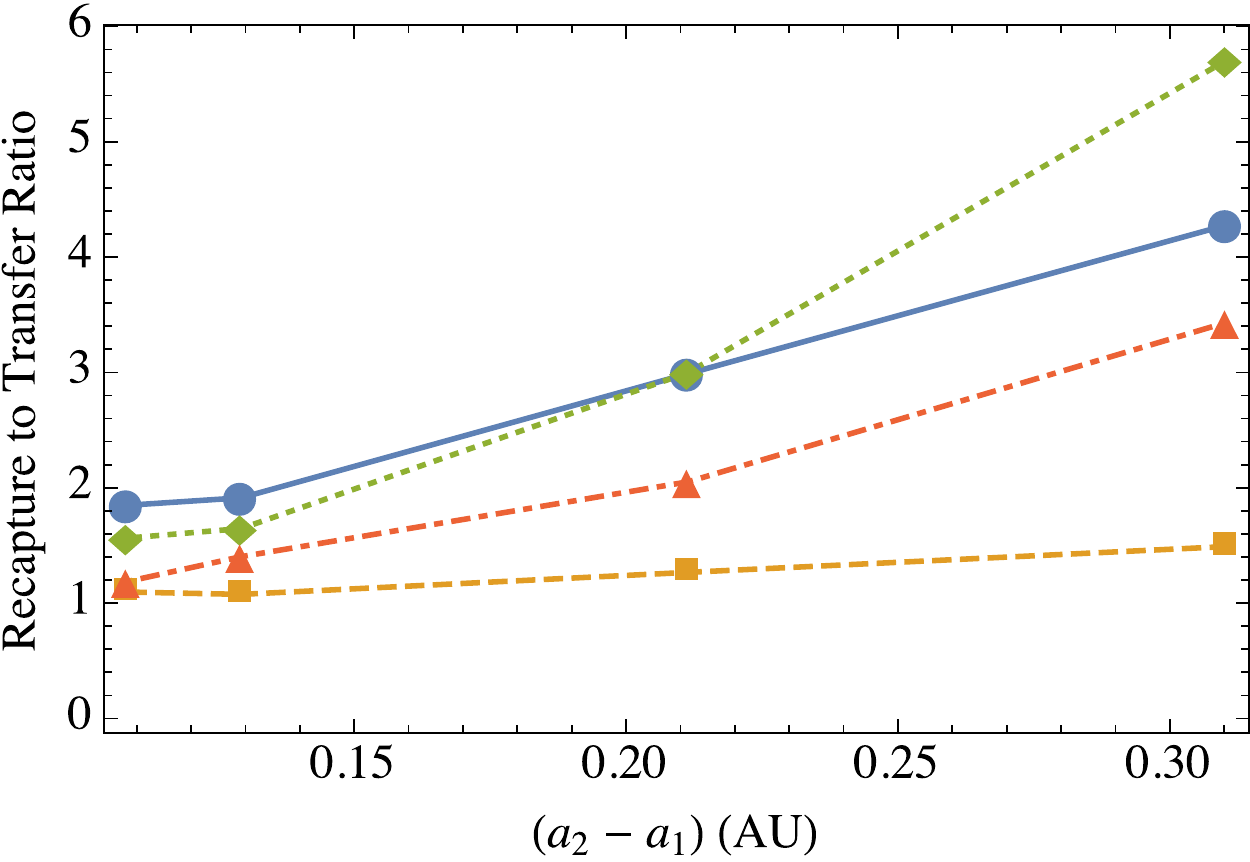}
\caption{Ratio of recaptured partices to transferred particles as a function of the difference in semi-major axis.  The blue circles (solid line) are when the inner planet is the source using the standard velocity distribution.  The orange squares (dashed) is for standard velocities when the outer planet is the source.  The green diamonds (dotted) and red triangles (dot-dashed) are for the restricted velocity distribution for the inner and outer planets, respectively, as the source.}\label{recaptrans}
\end{figure}

Another issue that we investigate with our simulations is the ratio of the ejecta that is recaptured on the source planet to the ejecta that is successfully transferred to the other planet.  These results are shown in Figure \ref{recaptrans}.  There we see that the recapture rate is similar to the transfer rate---the ratio being between unity and a few.  The ratio is within a factor of two when the planets are near the 6:5 or 7:6 MMR.  This fact implies that for multihabitable systems, biological transfer from one planet to another is similar in importance to auto-panspermia where biological material is transferred across the surface of a planet by its own collision ejecta---at least that portion of the ejecta that initially escapes the planet.

\subsection{Effects of Caustic Flows}

\begin{figure}
\includegraphics[width=0.49\textwidth]{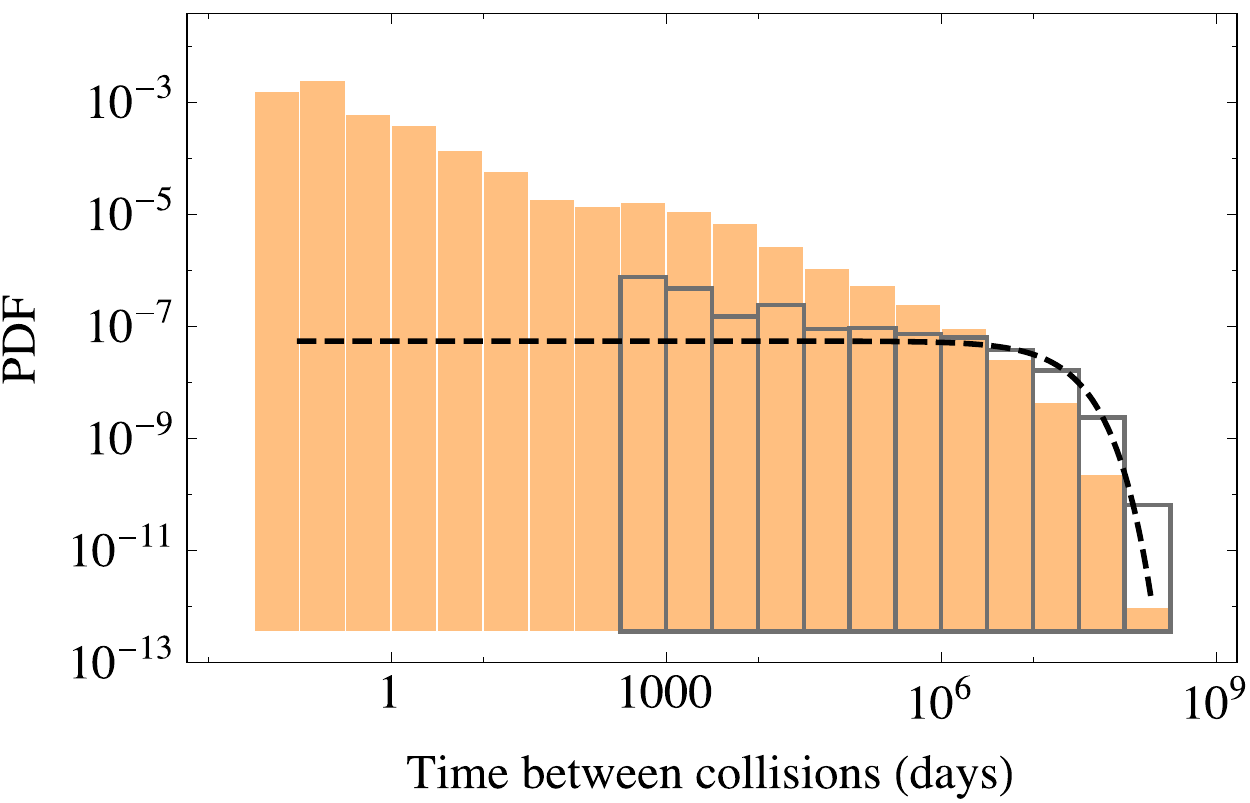}
\includegraphics[width=0.49\textwidth]{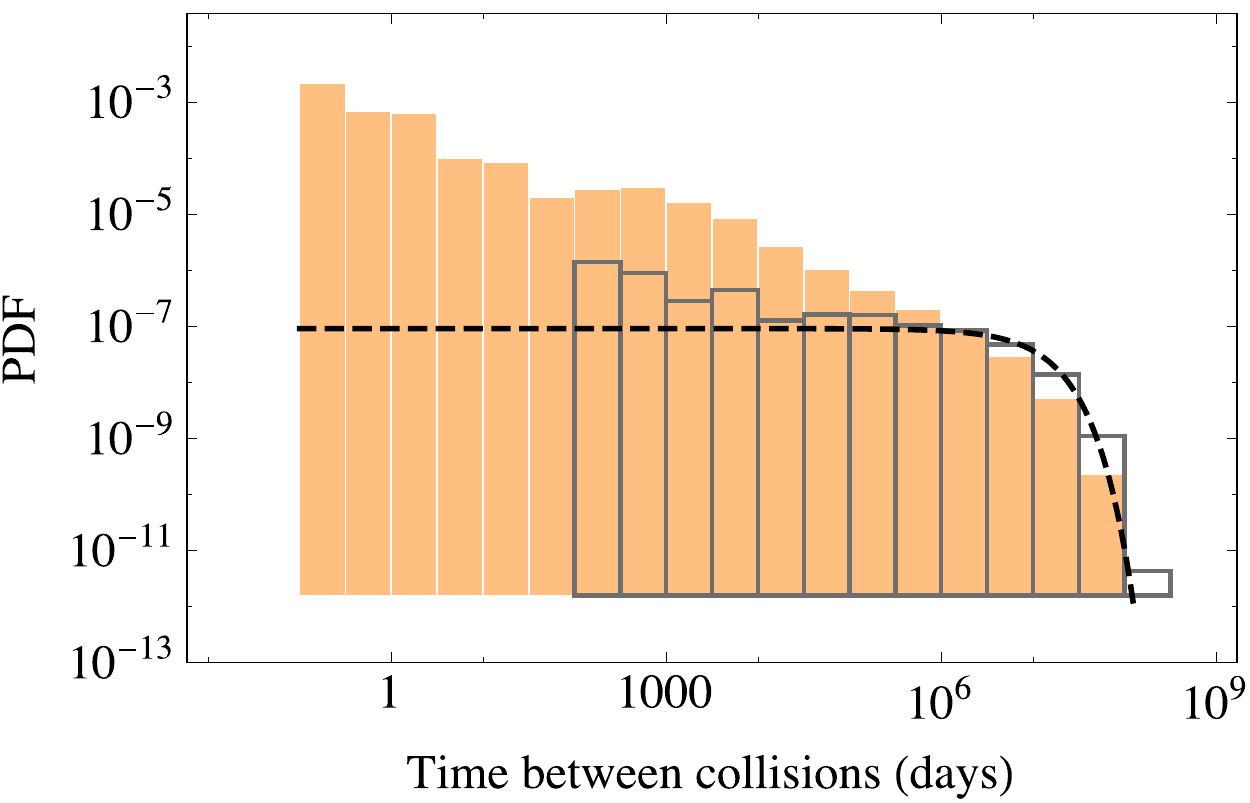}
\caption{Top: PDF of time differences between impacts for the 3:2 MMR pair, where the outer planet is the source, using our standard velocity distribution (outline) and using the restricted distribution (orange, solid).  The dashed line corresponds to the best fit exponential distribution to the standard velocity case.  Bottom: Similar plot but for the 6:5 MMR pair and with the inner planet being the source.  The excess of collisions at short time intervals in both of these plots shows the non-Poisson nature of the impacts.  Specifically, many collisions occur in quick succession---a result consistent with the expected behavior from caustic flows.   Results for all other MMR pairs and both sources are qualitatively similar to these.}\label{waittime}
\end{figure}

One of our earlier claims is that there should be an excess of successive collisions separated by only a short time interval.  Figure \ref{waittime} shows the distribution of separation times between successive collisions for two example cases---the non-resonant 3:2 pair using the outer planet as the source and the 6:5 pair using the inner planet as the source.  We examine results from both the standard velocity distribution and the restricted distribution.  Also shown is the best-fitting exponential distribution to the standard velocity case.

For the most part, near the long-separation tail of the distribution, the time differences do follow an exponential distribution---showing that a portion of the time distribution follows from Poisson fluctuations.  At short separations there is a slight to significant excess, indicating a non-Poisson component consistent with caustic flows.  The simulations that used the restricted velocity distribution show a very large excess of short time intervals between collisions.  This result is due to the much higher concentration of the particle velocities---giving a phase space that is much more densely filled.  Such flows would imply that multiple successful transfers are likely to occur in quick succession, spreading any biological material---which came from the same location of the source planet---across a range of locations on the destination planet.

\section{Discussion}\label{discussion}

Given our results above, we offer a few observations about life in multihabitable systems.  Both issues considered here turned out to be favorable for life and its transfer among multiple bodies in the nominal habitable zone of a multihabitable system.  Only in special situations do the orbital obliquity of the close-proximity planets vary by large amounts.  Simultaneously, the probability of material being shared between the planets is much larger than it is for the solar system.

The different dependencies that we observed in our simulations (especially the preference for low ejection velocities) bodes well for the successful transfer biological material for several reasons.  First, since smaller velocity particles are more readily transferred to the other planet, the collision that produces the ejecta can be less energetic---reducing the risk of destroying important material.  Second, lower particle velocities mean that more of the surface of the planet is effective as a source for the ejecta (high velocity particles had large changes in success rates that varied with longitude while low velocity particles were uniformly distributed in longitude).

Third, since the ejecta forms caustics with phase-space sheets having high particle densities, the successful transfer of one particle implies an increased probability that additional particles will also transfer successfully and within a relatively short amount of time.  Thus, one can imagine material from one part of the source planet being distributed almost simultaneously across much of the surface of the destination planet---giving the seeds of life a greater chance of taking hold.  The same effect is less likely to occur in more widely separated planet planetary systems (such as in the solar system), which rely on secular resonances to excite orbital eccentricities, because chaotic behavior near those resonances would, over long timescales, cause neighboring orbits to diverge.

Not only will panspermia be more common in a multihabitable system than in the solar system, but the close proximity of the planets to each other within the habitable zone of the host star allows for a real possibility of the planets having regions of similar climate---perhaps allowing the microbiological family tree to extend across the system.  There are many things to consider in multihabitable systems, especially in the cases where intelligent life emerges.

For panspermia in a multihabitable system, the transfer rate of ejecta from one planet to the other is comparable to the rate of auto-panspermia.  In all of our simulations the rate at which material is recaptured by the source planet is a factor of one to a few times the rate that material transfers between the two planets (see Figure \ref{recaptrans}) but the ratio is not 10 or 100.  Thus, in these systems one may find a significant amount of exchange between the two planet surfaces.

\section{Conclusions}\label{conclusion}

The prospect of planetary systems containing multiple habitable planets is intriguing.  And, given some of the results of the \kepler\ mission, we may expect that such systems exist.  As a full census of habitable-zone planets comes ever closer to reality, we may have the opportunity to study one or a few of these systems in some detail.  There are many issues that have an impact on life in a planetary system.  Here we have only considered the effects of large obliquity variations due to strong dynamical interactions---which could hamper or preclude the development of intelligent life due to its associated changes in climate---and the possibility of frequent lithopanspermia in each system.  We note that our models did not consider possible effects of tides, spin-orbit resonances, or the potentially strong effect of inclination resonances in the systems.

We found that obliquity variations are generally not affected by the close proximity of the planets in a multihabitable system.  Also, obliquity variations of close pairs embedded in the solar system or of potentially habitable planets in a system with a close pair were not sufficient to significantly reduce the probability of having a stable climate.  Only in cases where the inclination modal frequencies coincide with the planetary precession frequency did large obliquity variation arise.

At the same time, we expect to find relatively high rates of panspermia in these systems.  The nearer the planets are to one another the higher the success rate of ejecta transfer---coming quite close to the same rate as ejecta falling back to the surface of the originating planet.  Also, we found that the smaller the velocity of the ejected material the more uniformly they can be sourced across the originating planet.  With high velocity ejecta, the range of initial longitudes is constrained relative to the direction of motion.

Mean-motion resonance did not have as strong of an effect on the collision rate as the variations that occur from changes to the initial orbital parameters of the planets.  By way of comparison, changes to the initial conditions of the collision ejecta (within the parameters of the velocity distribution) produced variations consistent with Poisson fluctuations over the 10 million year integration time.  That is, for a given set of planet initial conditions, the transfer rate was robust given our simulation method.

Finally, we claimed that the restricted phase space of the ejected particles (due to their originating from a highly localized source and their collisionless nature) implies that the particles should form caustic flows of high density sheets of ejecta.  A consequence of such a dynamical state is non-Poisson fluctuations in the transfer rate occurring at short time intervals.  We observe these fluctuations, which are especially pronounced for the restricted velocity dispersion where the phase space is much more densely sampled by our particles.

\section*{Acknowledgements}
We thank Konstantin Batygin for many useful discussions in the preparation of this work.  J.H.S. acknowledges support from the Lindheimer Fellowship at CIERA, Northwestern and from NASA under grant NNX08AR04G issued through the Kepler Participating Scientist Program.


\bibliographystyle{plainnat}
\bibliography{msref}

\begin{thebibliography}{57}
\providecommand{\natexlab}[1]{#1}
\providecommand{\url}[1]{\texttt{#1}}
\expandafter\ifx\csname urlstyle\endcsname\relax
  \providecommand{\doi}[1]{doi: #1}\else
  \providecommand{\doi}{doi: \begingroup \urlstyle{rm}\Url}\fi

\bibitem[{Armstrong} et~al.(2004){Armstrong}, {Leovy}, and
  {Quinn}]{Armstrong:2004}
J.~C. {Armstrong}, C.~B. {Leovy}, and T.~{Quinn}.
\newblock {A 1 Gyr climate model for Mars: new orbital statistics and the
  importance of seasonally resolved polar processes}.
\newblock \emph{\icarus}, 171:\penalty0 255--271, October 2004.
\newblock \doi{10.1016/j.icarus.2004.05.007}.

\bibitem[{Armstrong} et~al.(2014){Armstrong}, {Barnes}, {Domagal-Goldman},
  {Breiner}, {Quinn}, and {Meadows}]{Armstrong:2014}
J.~C. {Armstrong}, R.~{Barnes}, S.~{Domagal-Goldman}, J.~{Breiner}, T.~R.
  {Quinn}, and V.~S. {Meadows}.
\newblock {Effects of Extreme Obliquity Variations on the Habitability of
  Exoplanets}.
\newblock \emph{Astrobiology}, 14:\penalty0 277--291, April 2014.
\newblock \doi{10.1089/ast.2013.1129}.

\bibitem[{Batalha} et~al.(2013)]{Batalha:2013}
N.~M. {Batalha} et~al.
\newblock {Planetary Candidates Observed by Kepler. III. Analysis of the First
  16 Months of Data}.
\newblock \emph{\apjs}, 204:\penalty0 24, February 2013.
\newblock \doi{10.1088/0067-0049/204/2/24}.

\bibitem[{Batygin} and {Brown}(2010)]{Batygin10}
K.~{Batygin} and M.~E. {Brown}.
\newblock {Early Dynamical Evolution of the Solar System: Pinning Down the
  Initial Conditions of the Nice Model}.
\newblock \emph{\apj}, 716:\penalty0 1323--1331, June 2010.
\newblock \doi{10.1088/0004-637X/716/2/1323}.

\bibitem[{Belbruno} et~al.(2012){Belbruno}, {Moro-Mart{\'{\i}}n}, {Malhotra},
  and {Savransky}]{Belbruno:2012}
E.~{Belbruno}, A.~{Moro-Mart{\'{\i}}n}, R.~{Malhotra}, and D.~{Savransky}.
\newblock {Chaotic Exchange of Solid Material Between Planetary Systems:
  Implications for Lithopanspermia}.
\newblock \emph{Astrobiology}, 12:\penalty0 754--774, August 2012.
\newblock \doi{10.1089/ast.2012.0825}.

\bibitem[{Borucki} et~al.(2010)]{Borucki:2010}
W.~J. {Borucki} et~al.
\newblock {Kepler Planet-Detection Mission: Introduction and First Results}.
\newblock \emph{Science}, 327:\penalty0 977--, February 2010.
\newblock \doi{10.1126/science.1185402}.

\bibitem[{Brasser} and {Walsh}(2011)]{Brasser11}
R.~{Brasser} and K.~J. {Walsh}.
\newblock {Stability analysis of the martian obliquity during the Noachian
  era}.
\newblock \emph{\icarus}, 213:\penalty0 423--427, May 2011.
\newblock \doi{10.1016/j.icarus.2011.02.024}.

\bibitem[{Brasser} et~al.(2014){Brasser}, {Ida}, and {Kokubo}]{Brasser:2014}
R.~{Brasser}, S.~{Ida}, and E.~{Kokubo}.
\newblock {A dynamical study on the habitability of terrestrial exoplanets - II
  The super-Earth HD 40307 g}.
\newblock \emph{\mnras}, 440:\penalty0 3685--3700, June 2014.
\newblock \doi{10.1093/mnras/stu555}.

\bibitem[{Burke} et~al.(2014)]{Burke:2014}
C.~J. {Burke} et~al.
\newblock {Planetary Candidates Observed by Kepler IV: Planet Sample from Q1-Q8
  (22 Months)}.
\newblock \emph{\apjs}, 210:\penalty0 19, February 2014.
\newblock \doi{10.1088/0067-0049/210/2/19}.

\bibitem[{Carter} et~al.(2012){Carter}, {Agol}, et~al.]{Carter:2012}
J.~A. {Carter}, E.~{Agol}, et~al.
\newblock {Kepler-36: A Pair of Planets with Neighboring Orbits and Dissimilar
  Densities}.
\newblock \emph{Science}, 337:\penalty0 556--, August 2012.
\newblock \doi{10.1126/science.1223269}.

\bibitem[{Chambers}(1999)]{Chambers:1999}
J.~E. {Chambers}.
\newblock {A hybrid symplectic integrator that permits close encounters between
  massive bodies}.
\newblock \emph{\mnras}, 304:\penalty0 793--799, April 1999.
\newblock \doi{10.1046/j.1365-8711.1999.02379.x}.

\bibitem[{Chandler} and {Sohl}(2000)]{Chandler00}
M.~A. {Chandler} and L.~E. {Sohl}.
\newblock {Climate forcings and the initiation of low-latitude ice sheets
  during the Neoproterozoic Varanger glacial interval}.
\newblock \emph{\jgr}, 105:\penalty0 20737--20756, 2000.
\newblock \doi{10.1029/2000JD900221}.

\bibitem[{Colombo}(1966)]{Colombo66}
G.~{Colombo}.
\newblock {Cassini's second and third laws}.
\newblock \emph{\aj}, 71:\penalty0 891, November 1966.
\newblock \doi{10.1086/109983}.

\bibitem[{Deck} et~al.(2012){Deck}, {Holman}, {Agol}, {Carter}, {Lissauer},
  {Ragozzine}, and {Winn}]{Deck:2012}
K.~M. {Deck}, M.~J. {Holman}, E.~{Agol}, J.~A. {Carter}, J.~J. {Lissauer},
  D.~{Ragozzine}, and J.~N. {Winn}.
\newblock {Rapid Dynamical Chaos in an Exoplanetary System}.
\newblock \emph{\apjl}, 755:\penalty0 L21, August 2012.
\newblock \doi{10.1088/2041-8205/755/1/L21}.

\bibitem[{Dones} et~al.(1999){Dones}, {Gladman}, {Melosh}, {Tonks}, {Levison},
  and {Duncan}]{Dones:1999}
L.~{Dones}, B.~{Gladman}, H.~J. {Melosh}, W.~B. {Tonks}, H.~F. {Levison}, and
  M.~{Duncan}.
\newblock {Dynamical Lifetimes and Final Fates of Small Bodies: Orbit
  Integrations vs {\"O}pik Calculations}.
\newblock \emph{\icarus}, 142:\penalty0 509--524, December 1999.
\newblock \doi{10.1006/icar.1999.6220}.

\bibitem[{Doyle} et~al.(2011)]{Doyle:2011}
L.~R. {Doyle} et~al.
\newblock {Kepler-16: A Transiting Circumbinary Planet}.
\newblock \emph{Science}, 333:\penalty0 1602--, September 2011.
\newblock \doi{10.1126/science.1210923}.

\bibitem[{Fanale} et~al.(1982){Fanale}, {Salvail}, {Banerdt}, and
  {Saunders}]{Fanale82}
F.~P. {Fanale}, J.~R. {Salvail}, W.~B. {Banerdt}, and R.~S. {Saunders}.
\newblock {Mars - The regolith-atmosphere-cap system and climate change}.
\newblock \emph{\icarus}, 50:\penalty0 381--407, June 1982.
\newblock \doi{10.1016/0019-1035(82)90131-2}.

\bibitem[{Francois} et~al.(1990){Francois}, {Walker}, and {Kuhn}]{Francois90}
L.~M. {Francois}, J.~C.~G. {Walker}, and W.~R. {Kuhn}.
\newblock {A numerical simulation of climate changes during the obliquity cycle
  on Mars}.
\newblock \emph{\jgr}, 95:\penalty0 14761--14778, August 1990.
\newblock \doi{10.1029/JB095iB09p14761}.

\bibitem[{Gladman} et~al.(1996){Gladman}, {Burns}, {Duncan}, {Lee}, and
  {Levison}]{Gladman:1996}
B.~J. {Gladman}, J.~A. {Burns}, M.~{Duncan}, P.~{Lee}, and H.~F. {Levison}.
\newblock {The Exchange of Impact Ejecta Between Terrestrial Planets}.
\newblock \emph{Science}, 271:\penalty0 1387--1392, March 1996.
\newblock \doi{10.1126/science.271.5254.1387}.

\bibitem[{G{\'o}rski} et~al.(2005){G{\'o}rski}, {Hivon}, {Banday}, {Wandelt},
  {Hansen}, {Reinecke}, and {Bartelmann}]{Gorski:2005}
K.~M. {G{\'o}rski}, E.~{Hivon}, A.~J. {Banday}, B.~D. {Wandelt}, F.~K.
  {Hansen}, M.~{Reinecke}, and M.~{Bartelmann}.
\newblock {HEALPix: A Framework for High-Resolution Discretization and Fast
  Analysis of Data Distributed on the Sphere}.
\newblock \emph{\apj}, 622:\penalty0 759--771, April 2005.
\newblock \doi{10.1086/427976}.

\bibitem[{Hays} et~al.(1976){Hays}, {Imbrie}, and {Shackleton}]{Hays76}
J.~D. {Hays}, J.~{Imbrie}, and N.~J. {Shackleton}.
\newblock {Variations in the Earth's Orbit: Pacemaker of the Ice Ages}.
\newblock \emph{Science}, 194:\penalty0 1121--1132, December 1976.
\newblock \doi{10.1126/science.194.4270.1121}.

\bibitem[{Imbrie}(1982)]{Imbrie82}
J.~{Imbrie}.
\newblock {Astronomical theory of the Pleistocene ice ages - A brief historical
  review}.
\newblock \emph{\icarus}, 50:\penalty0 408--422, June 1982.
\newblock \doi{10.1016/0019-1035(82)90132-4}.

\bibitem[{Jenkins}(2000)]{Jenkins00}
G.~S. {Jenkins}.
\newblock {Global climate model high-obliquity solutions to the ancient climate
  puzzles of the Faint-Young Sun Paradox and low-latitude Proterozoic
  Glaciation}.
\newblock \emph{\jgr}, 105:\penalty0 7357--7370, 2000.
\newblock \doi{10.1029/1999JD901125}.

\bibitem[{Kinoshita}(1972)]{Kinoshita:1972}
H.~{Kinoshita}.
\newblock {First-Order Perturbations of the Two Finite Body Problem}.
\newblock \emph{\pasj}, 24:\penalty0 423, 1972.

\bibitem[{Laskar}(1996)]{Laskar96}
J.~{Laskar}.
\newblock {Large Scale Chaos and Marginal Stability in the Solar System}.
\newblock \emph{Celestial Mechanics and Dynamical Astronomy}, 64:\penalty0
  115--162, March 1996.
\newblock \doi{10.1007/BF00051610}.

\bibitem[{Laskar} and {Robutel}(1993)]{Laskar93b}
J.~{Laskar} and P.~{Robutel}.
\newblock {The chaotic obliquity of the planets}.
\newblock \emph{\nat}, 361:\penalty0 608--612, February 1993.
\newblock \doi{10.1038/361608a0}.

\bibitem[{Laskar} et~al.(1993){Laskar}, {Joutel}, and {Robutel}]{Laskar93a}
J.~{Laskar}, F.~{Joutel}, and P.~{Robutel}.
\newblock {Stabilization of the earth's obliquity by the moon}.
\newblock \emph{\nat}, 361:\penalty0 615--617, February 1993.
\newblock \doi{10.1038/361615a0}.

\bibitem[{Lee} and {Peale}(2002)]{Lee02}
M.~H. {Lee} and S.~J. {Peale}.
\newblock {Dynamics and Origin of the 2:1 Orbital Resonances of the GJ 876
  Planets}.
\newblock \emph{\apj}, 567:\penalty0 596--609, March 2002.
\newblock \doi{10.1086/338504}.

\bibitem[{Li} and {Batygin}(2014{\natexlab{a}})]{Li14a}
G.~{Li} and K.~{Batygin}.
\newblock {On the Spin-axis Dynamics of a Moonless Earth}.
\newblock \emph{\apj}, 790:\penalty0 69, July 2014{\natexlab{a}}.
\newblock \doi{10.1088/0004-637X/790/1/69}.

\bibitem[{Li} and {Batygin}(2014{\natexlab{b}})]{Li14b}
G.~{Li} and K.~{Batygin}.
\newblock {Pre-late Heavy Bombardment Evolution of the Earth's Obliquity}.
\newblock \emph{\apj}, 795:\penalty0 67, November 2014{\natexlab{b}}.
\newblock \doi{10.1088/0004-637X/795/1/67}.

\bibitem[{Lissauer} et~al.(2012){Lissauer}, {Barnes}, and
  {Chambers}]{Lissauer12}
J.~J. {Lissauer}, J.~W. {Barnes}, and J.~E. {Chambers}.
\newblock {Obliquity variations of a moonless Earth}.
\newblock \emph{\icarus}, 217:\penalty0 77--87, January 2012.
\newblock \doi{10.1016/j.icarus.2011.10.013}.

\bibitem[{Lissauer} et~al.(2011)]{Lissauer:2011b}
J.~J. {Lissauer} et~al.
\newblock {Architecture and Dynamics of Kepler's Candidate Multiple Transiting
  Planet Systems}.
\newblock \emph{\apjs}, 197:\penalty0 8, November 2011.
\newblock \doi{10.1088/0067-0049/197/1/8}.

\bibitem[{Lissauer} et~al.(2014)]{Lissauer:2014}
J.~J. {Lissauer} et~al.
\newblock {Validation of Kepler's Multiple Planet Candidates. II. Refined
  Statistical Framework and Descriptions of Systems of Special Interest}.
\newblock \emph{\apj}, 784:\penalty0 44, March 2014.
\newblock \doi{10.1088/0004-637X/784/1/44}.

\bibitem[{McKay} et~al.(1996){McKay}, {Gibson}, {Thomas-Keprta}, {Vali},
  {Romanek}, {Clemett}, {Chillier}, {Maechling}, and {Zare}]{McKay:1996}
D.~S. {McKay}, E.~K. {Gibson}, Jr., K.~L. {Thomas-Keprta}, H.~{Vali}, C.~S.
  {Romanek}, S.~J. {Clemett}, X.~D.~F. {Chillier}, C.~R. {Maechling}, and R.~N.
  {Zare}.
\newblock {Search for Past Life on Mars: Possible Relic Biogenic Activity in
  Martian Meteorite ALH84001}.
\newblock \emph{Science}, 273:\penalty0 924--930, August 1996.
\newblock \doi{10.1126/science.273.5277.924}.

\bibitem[{Melosh}(1988)]{Melosh:1998}
H.~J. {Melosh}.
\newblock {The rocky road to panspermia}.
\newblock \emph{\nat}, 332:\penalty0 687--688, April 1988.
\newblock \doi{10.1038/332687a0}.

\bibitem[{Melosh}(2003)]{Melosh:2003}
H.~J. {Melosh}.
\newblock {Exchange of Meteorites (and Life?) Between Stellar Systems}.
\newblock \emph{Astrobiology}, 3:\penalty0 207--215, January 2003.
\newblock \doi{10.1089/153110703321632525}.

\bibitem[{Melosh} and {Tonks}(1993)]{Melosh:1993}
H.~J. {Melosh} and W.~B. {Tonks}.
\newblock {Swapping Rocks: Ejection and Exchange of Surface Material Among the
  Terrestrial Planets}.
\newblock \emph{Meteoritics}, 28:\penalty0 398--398, July 1993.

\bibitem[{Mileikowsky} et~al.(2000){Mileikowsky}, {Cucinotta}, {Wilson},
  {Gladman}, {Horneck}, {Lindegren}, {Melosh}, {Rickman}, {Valtonen}, and
  {Zheng}]{Mileikowsky:2000}
C.~{Mileikowsky}, F.~A. {Cucinotta}, J.~W. {Wilson}, B.~{Gladman},
  G.~{Horneck}, L.~{Lindegren}, J.~{Melosh}, H.~{Rickman}, M.~{Valtonen}, and
  J.~Q. {Zheng}.
\newblock {Natural Transfer of Viable Microbes in Space. 1. From Mars to Earth
  and Earth to Mars}.
\newblock \emph{\icarus}, 145:\penalty0 391--427, June 2000.
\newblock \doi{10.1006/icar.1999.6317}.

\bibitem[{Mullally} et~al.(2015)]{Mullaly:2015}
F.~{Mullally} et~al.
\newblock {Planetary Candidates Observed by Kepler VI: Planet Sample from
  Q1-Q16 (47 Months)}.
\newblock \emph{ArXiv e-prints:1502.02038}, February 2015.

\bibitem[{Nakamura} and {Tajika}(2003)]{Nakamura03}
T.~{Nakamura} and E.~{Tajika}.
\newblock {Climate change of Mars-like planets due to obliquity variations:
  implications for Mars}.
\newblock \emph{\grl}, 30:\penalty0 1685, July 2003.
\newblock \doi{10.1029/2002GL016725}.

\bibitem[{Neron de Surgy} and {Laskar}(1997)]{deSurgy97}
O.~{Neron de Surgy} and J.~{Laskar}.
\newblock {On the long term evolution of the spin of the Earth.}
\newblock \emph{\aap}, 318:\penalty0 975--989, February 1997.

\bibitem[{Pollack} and {Toon}(1982)]{Pollack82}
J.~B. {Pollack} and O.~B. {Toon}.
\newblock {Quasi-periodic climate changes on Mars - A review}.
\newblock \emph{\icarus}, 50:\penalty0 259--287, June 1982.
\newblock \doi{10.1016/0019-1035(82)90126-9}.

\bibitem[{Rowe} et~al.(2015)]{Rowe:2015}
J.~F. {Rowe} et~al.
\newblock {Planetary Candidates Observed by Kepler. V. Planet Sample from
  Q1-Q12 (36 Months)}.
\newblock \emph{\apjs}, 217:\penalty0 16, March 2015.
\newblock \doi{10.1088/0067-0049/217/1/16}.

\bibitem[{Sikivie}(1998)]{Sikivie:1998}
P.~{Sikivie}.
\newblock {Caustic rings of dark matter}.
\newblock \emph{Physics Letters B}, 432:\penalty0 139--144, July 1998.
\newblock \doi{10.1016/S0370-2693(98)00595-4}.

\bibitem[{Soto} et~al.(2012){Soto}, {Mischna}, and {Richardson}]{Soto12}
A.~{Soto}, M.~A. {Mischna}, and M.~I. {Richardson}.
\newblock {Climate Dynamics of Atmospheric Collapse on Ancient Mars}.
\newblock In \emph{Lunar and Planetary Institute Science Conference Abstracts},
  volume~43 of \emph{Lunar and Planetary Institute Science Conference
  Abstracts}, page 2783, March 2012.

\bibitem[{Spiegel} et~al.(2009){Spiegel}, {Menou}, and {Scharf}]{Spiegel09}
D.~S. {Spiegel}, K.~{Menou}, and C.~A. {Scharf}.
\newblock {Habitable Climates: The Influence of Obliquity}.
\newblock \emph{\apj}, 691:\penalty0 596--610, January 2009.
\newblock \doi{10.1088/0004-637X/691/1/596}.

\bibitem[{Steffen} et~al.(2010)]{Steffen:2010}
J.~H. {Steffen} et~al.
\newblock {Five Kepler Target Stars That Show Multiple Transiting Exoplanet
  Candidates}.
\newblock \emph{\apj}, 725:\penalty0 1226--1241, December 2010.
\newblock \doi{10.1088/0004-637X/725/1/1226}.

\bibitem[{Toon} et~al.(1980){Toon}, {Pollack}, {Ward}, {Burns}, and
  {Bilski}]{Toon80}
O.~B. {Toon}, J.~B. {Pollack}, W.~{Ward}, J.~A. {Burns}, and K.~{Bilski}.
\newblock {The astronomical theory of climatic change on Mars}.
\newblock \emph{\icarus}, 44:\penalty0 552--607, December 1980.
\newblock \doi{10.1016/0019-1035(80)90130-X}.

\bibitem[{Touma} and {Wisdom}(1993)]{Touma93}
J.~{Touma} and J.~{Wisdom}.
\newblock {The chaotic obliquity of Mars}.
\newblock \emph{Science}, 259:\penalty0 1294--1297, February 1993.
\newblock \doi{10.1126/science.259.5099.1294}.

\bibitem[{Vernekar}(1972)]{Vernekar72}
A.~D. {Vernekar}.
\newblock {a Study of Mean Temperature of the Earth's Surface}.
\newblock In \emph{Atmospheric Radiation}, page 228, 1972.

\bibitem[{Ward}(1973)]{Ward73}
W.~R. {Ward}.
\newblock {Large-Scale Variations in the Obliquity of Mars}.
\newblock \emph{Science}, 181:\penalty0 260--262, July 1973.
\newblock \doi{10.1126/science.181.4096.260}.

\bibitem[{Weertman}(1976)]{Weertman76}
J.~{Weertman}.
\newblock {Milankovitch solar radiation variations and ice age ice sheet
  sizes}.
\newblock \emph{\nat}, 261:\penalty0 17--20, May 1976.
\newblock \doi{10.1038/261017a0}.

\bibitem[{Wells} et~al.(2003){Wells}, {Armstrong}, and {Gonzalez}]{Wells:2003}
L.~E. {Wells}, J.~C. {Armstrong}, and G.~{Gonzalez}.
\newblock {Reseeding of early earth by impacts of returning ejecta during the
  late heavy bombardment}.
\newblock \emph{\icarus}, 162:\penalty0 38--46, March 2003.
\newblock \doi{10.1016/S0019-1035(02)00077-5}.

\bibitem[{Welsh} et~al.(2012)]{Welsh:2012b}
W.~F. {Welsh} et~al.
\newblock {Transiting circumbinary planets Kepler-34 b and Kepler-35 b}.
\newblock \emph{\nat}, 481:\penalty0 475--479, January 2012.
\newblock \doi{10.1038/nature10768}.

\bibitem[{Williams} and {Kasting}(1997)]{Williams97}
D.~M. {Williams} and J.~F. {Kasting}.
\newblock {Habitable Planets with High Obliquities}.
\newblock \emph{\icarus}, 129:\penalty0 254--267, September 1997.
\newblock \doi{10.1006/icar.1997.5759}.

\bibitem[{Worth} et~al.(2013){Worth}, {Sigurdsson}, and {House}]{Worth:2013}
R.~J. {Worth}, S.~{Sigurdsson}, and C.~H. {House}.
\newblock {Seeding Life on the Moons of the Outer Planets via Lithopanspermia}.
\newblock \emph{Astrobiology}, 13:\penalty0 1155--1165, December 2013.
\newblock \doi{10.1089/ast.2013.1028}.

\bibitem[{Xie}(2013)]{Xie:2013}
J.-W. {Xie}.
\newblock {Transit Timing Variation of Near-resonance Planetary Pairs:
  Confirmation of 12 Multiple-planet Systems}.
\newblock \emph{\apjs}, 208:\penalty0 22, October 2013.
\newblock \doi{10.1088/0067-0049/208/2/22}.

\end{thebibliography}

\appendix

\section{Hamiltonian Equations}

The Hamiltonian we use for our simulations \citep[e.g.][]{Colombo66, Kinoshita:1972, Laskar93a, Touma93, deSurgy97,Armstrong:2004} is given by:
\begin{eqnarray}
\label{eqn:HF}
H(\chi, \psi, t) &=& \frac{1}{2}\alpha\chi^2 +\sqrt{1-\chi^2} \left( A(t)\sin{\psi}+B(t)\cos{\psi}) \right),
\end{eqnarray}
where $\chi = \cos \epsilon$, $\epsilon$ is the obliquity, $\psi$ is the longitude of the spin axis, $\alpha$ is the precession coefficient defined as \citep{deSurgy97}, 
\begin{eqnarray}
\label{e:alpha}
\alpha &=& \frac{3G}{2\omega}\frac{m_*}{(a\sqrt{1-e^2})^3} E_d ,
\end{eqnarray}
and 
\begin{eqnarray}
A(t) = 2(\dot{q}+p(q\dot{p}-p\dot{q}))/\sqrt{1-p^2-q^2}, \\
B(t) = 2(\dot{p}-q(q\dot{p}-p\dot{q}))/\sqrt{1-p^2-q^2},
\end{eqnarray}
where $p=\sin{i/2}~\sin{\Omega}$ and $q=\sin{i/2}~\cos{\Omega}$. $A(t)$ and $B(t)$ are obtained using the aforementioned results of the orbital elements from our MERCURY simulations.  In the expression for $\alpha$, $m_*$ is the mass of the star, $a$ and $e$ are the semi-major axis and the eccentricity of the planet's orbit, $\omega$ is the spin rate of the planet, $\Omega$ is the longitude of ascending node, and $E_d$ is the dynamical ellipticity of the planet.

\section{Initial conditions}

Table \ref{initcond} shows the initial orbital elements we used for our simulations.  Prior to including the test particle ejecta we transformed the positions and velocities such that the orbital distance of the inner planet was 1 AU.  In all simulations, the mass of the central star is 1 $M_\odot$ and the mass of the planets are $1.0\times 10^{-6} M_\odot$ and the planet densities are 1 g/cc.

\begin{table}[!h]
\caption{Simulation initial conditions.}
\label{initcond}
\begin{tabular}{lrrrrrrr} \hline
Run & Planet & $a$ & $e$ & $i$ & $\varpi$ & $\Omega$ & $M$ \\ \hline \hline
3:2 resonant          & 1 & 0.996185E+00 & 0.217249E-01 & 0.997871E+00 & 0.346938E+03 & 0.301425E+00 & 0.162046E+03 \\
                             & 2 & 0.130541E+01 & 0.234200E-01 & 0.700569E+00 & 0.165285E+03 & 0.359627E+03 & 0.500250E+02 \\
3:2 nonresonant N & 1 & 0.996185E+00 & 0.217249E-01 & 0.997871E+00 & 0. & 0. & 0. \\
                             & 2 & 0.130541E+01 & 0.234200E-01 & 0.700569E+00 & 0. & 0. & 0. \\
3:2 nonresonant I  & 1 & 0.996185E+00 & 0.217249E-01 & 0.997871E+00 & 0. & 120. & 240. \\
                             & 2 & 0.130541E+01 & 0.234200E-01 & 0.700569E+00 & 0.165285E+03 & 0.359627E+03 & 0.500250E+02 \\
4:3 resonant          & 1 & 0.964821E+00 & 0.218355E-01 & 0.685909E+00 & 0.207332E+03 & 0.455775E-02 & 0.214100E+03 \\
                             & 2 & 0.116885E+01 & 0.223782E-01 & 0.983627E+00 & 0.125833E+02 & 0.359761E+03 & 0.354742E+03 \\
4:3 nonresonant N & 1 & 0.964821E+00 & 0.218355E-01 & 0.685909E+00 & 0. & 0. & 0. \\
                             & 2 & 0.116885E+01 & 0.223782E-01 & 0.983627E+00 & 0. & 0. & 0. \\
4:3 nonresonant I  & 1 & 0.964821E+00 & 0.218355E-01 & 0.685909E+00 & 0. & 120. & 240. \\
                             & 2 & 0.116885E+01 & 0.223782E-01 & 0.983627E+00 & 0.125833E+02 & 0.359761E+03 & 0.354742E+03 \\
6:5 resonant          & 1 & 0.953590E-01 & 0.199269E-01 & 0.983127E+00 & 0.232531E+02 & 0.445968E+01 & 0.177217E+03 \\
                             & 2 & 0.107684E+00 & 0.207574E-01 & 0.721238E+00 & 0.202503E+03 & 0.354280E+03 & 0.148131E+03 \\
6:5 nonresonant N & 1 & 0.953590E-01 & 0.199269E-01 & 0.983127E+00 & 0. & 0. & 0. \\
                             & 2 & 0.107684E+00 & 0.207574E-01 & 0.721238E+00 & 0. & 0. & 0. \\
7:6 resonant          & 1 & 0.976083E-01 & 0.182103E-01 & 0.994049E+00 & 0.128348E+03 & 0.358880E+03 & 0.105195E+03 \\
                             & 2 & 0.108174E+00 & 0.189198E-01 & 0.705235E+00 & 0.307745E+03 & 0.150213E+01 & 0.270803E+03 \\
7:6 nonresonant N & 1 & 0.976083E-01 & 0.182103E-01 & 0.994049E+00 & 0. & 0. & 0. \\
                             & 2 & 0.108174E+00 & 0.189198E-01 & 0.705235E+00 & 0. & 0. & 0. \\ \hline
\end{tabular}
\end{table}

\end{document}